\documentclass[a4paper]{article}
\usepackage{subcaption}
\usepackage{mathtools}
\usepackage{amsmath}
\usepackage{amssymb}
\usepackage{hyperref}
\usepackage{graphicx}
\usepackage{overpic}
\usepackage{braket}
\usepackage{appendix}
\usepackage{dsfont} 
\usepackage{bm}
\usepackage{tabularx}
\usepackage{amsthm}
\usepackage[english]{babel}
\usepackage{authblk}
\usepackage{cite}
\usepackage{siunitx} 
\usepackage{color} 
\usepackage{fouridx}

\newcommand{\beq}{\begin{equation}}
\newcommand{\eeq}{\end{equation}}
\newcommand{\eq}[1]{\begin{align}#1\end{align}}
\newcommand\Eq[1]{Eq.~(\ref{#1})}
\newcommand\Poincare{Poincar\'e }

\newcommand\inout{\text{in/out}}
\newcommand\out{\text{out}}
\newcommand\inn{\text{in}}

\DeclareMathOperator\arctantwo{atan2}

\DeclarePairedDelimiter\abs{\lvert}{\rvert}%
\title{The Polychromatic T-matrix}
\author[1]{Maxim Vavilin\footnote{maxim.vavilin@kit.edu}}
\author[2]{Ivan Fernandez-Corbaton}
\affil[1]{Institute of Theoretical Solid State Physics, Karlsruhe Institute of Technology, Karlsruhe, Germany}
\affil[2]{Institute of Nanotechnology, Karlsruhe Institute of Technology, Karlsruhe, Germany}
\date{}

\begin{document}
\maketitle

\pagenumbering{arabic}

\begin{abstract}
The T-matrix is a powerful tool that provides the complete description of the linear interaction between the electromagnetic field and a given object. In here, we generalize the usual monochromatic formalism to the case of polychromatic field-matter interaction. The group of transformations of special relativity provides the guidance for building the new formalism, which is inherently polychromatic. The polychromatic T-matrix affords the direct treatment of the interaction of electromagnetic pulses with objects, even when the objects move at constant relativistic speeds.
\end{abstract}

\tableofcontents

\section{Introduction}

Understanding and engineering the interaction between electromagnetic radiation and matter has been, is, and will be crucial for our scientific and technological development. These endeavors benefit from accurate an efficient theoretical and numerical tools of wide applicability. The T-matrix is a particularly useful formalism for the study of light-matter interactions featuring such beneficial properties. The T-matrix of an object is a linear operator that produces the field scattered by the object upon a given incident illumination. In the most common embodiment the incident field is expanded into regular multipolar fields, the scattered field is expanded into irregular outgoing multipolar fields, and the T-matrix connects the two sets of expansion coefficients. Following the seminal paper by Waterman \cite{waterman1965}, the T-matrix formalism has been established as one of the most powerful and popular techniques for computing the electromagnetic response of single and composite objects. The amount of research related to the T-matrix and its manifold applications grows at an increasing rate \cite{Gouesbet2019,Mishchenko2020}. 

For all its outstanding properties, the T-matrix formalism has still some limitations. One important question whose theoretical details and algorithmic answers are currently under intense study \cite{Theobald2017,Egel2017,Martin2019,Schebarchov2019,Lamprianidis2023} is the validity of the expansion of the scattered fields outside the object, but inside the smallest sphere circumscribing the object. Such question compromises the computation of the joint T-matrix of two objects that invade each others smallest circumscribing spheres. In here, we address another important limitation. The T-matrix formalism is monochromatic at its core, that is, it has been defined and developed systematically assuming that the illuminating fields are monochromatic, with time dependence $\exp(-i\omega_0 t)$ for some fixed frequency $\omega_0$. It is clear that the linearity of Maxwell equations permits the computation of the response of an object to an incident polychromatic field by superposing the responses to many monochromatic fields with different frequencies. Yet, as far as we know, a systematic development of the polychromatic T-matrix does not exist, which in particular leaves out the direct treatment of the interaction of objects with light pulses. Other cases where a robust polychromatic T-matrix formalism will be beneficial are linear processes that change the frequency of light. One example is the illumination of objects moving at constant speeds. While the source may be approximately monochromatic in its frame, new frequencies will appear after considering the light in the reference frame of the object. The problem is readily solved if the T-matrix in the rest frame of the object can be transformed with the appropriate Lorentz boost. Raman scattering is another example where the energy of internal vibrations in the object is added to or subtracted from the incident frequency, while the response is still linear from the point of view of the illumination.

We will use an approach to the T-matrix that is not in the main stream of T-matrix research, but that has already shown its value. The T-matrix can be approached from the perspective of group theory, that is, the study of symmetry transformations of physical systems \cite{tung1985}. Of special importance for the T-matrix is the concept of irreducible representations of a group of transformations, which can be understood as elementary components of e.g. an operator or a physical field, that transform in distinct ways under the transformations belonging to a given group, and that do not mix with each other upon such transformations. Starting from the T-matrix formalism that Waterman had developed for a single object \cite{waterman1965}, Peterson and Str\"om extended it to multiple objects with the help of group theory\cite{peterson1973}. They showed that the monochromatic T-matrix transforms according to the 3D Euclidean group, which consists of spatial translations and rotations, and exploited the transformation properties of the irreducible representations of such group for conveniently formulating the translations and rotations of individual T-matrices that are needed for computing the T-matrix of a composite object. The connection between the monochromatic T-matrix and the 3D Euclidean group is well illustrated by the fact that the translation theorems for vector spherical harmonics can be derived within the context of group theory \cite[Secs.~8.6 and 9.8]{tung1985}. 

The 3D Euclidean group is not sufficient for treating polychromatic electromagnetic fields. Wigner showed in a landmark paper \cite{wigner1939} that physical fields describing elementary particles such as electrons and photons transform under the Poincar\'e group, which consist of the 3D Euclidean group plus time translations and Lorentz boosts, also known as Lorentz transformations, or changes of inertial reference frame. Lorentz boosts change the frequency content of the field, as exemplified by the Doppler effect, which makes a monochromatic treatment impossible. The Poincar\'e group is the group of transformations of special relativity, and its relevance in electromagnetism can be appreciated in the decomposition of the electromagnetic field in irreducible parts \cite{moses1973}, and in the formulation of a scalar product \cite{gross1964} between two given electromagnetic fields whose result is invariant under all the transformations of the Poincar\'e group. Such scalar product enables one to use the tools of Hilbert spaces in electromagnetism and, in particular, allows one to systematically compute fundamental quantities such as energy and momentum contained in a given electromagnetic field \cite[\S 9, Chap.~3]{Birula1975}. Combining the scalar product with the S-matrix operator, one obtains a unified theory of conservation laws in light-matter interactions \cite{fernandez2017}. The S-matrix is a linear operator that maps irregular incoming fields to irregular outgoing fields, and that is numerically related to the T-matrix in a bijective and straightforward manner.

In this article, we extend the T-matrix formalism to the polychromatic setting by following the guidance provided by the representation of the Poincar\'e group of transformations in the Hilbert space of solutions of Maxwell equations. The formalism developed with the help of such algebraic structures and methods is inherently polychromatic, and facilitates the definitions of basis states and frequency integrals, among other aspects. The appropriate basis states and corresponding expansion coefficient functions for the fields are found by requiring that they transform unitarily according to the representations of the Poincar\'e group appropriate for photonic fields, as established by Wigner \cite{wigner1939}. The requirement of unitarity is fulfilled with respect to the scalar product whose result is invariant under all the transformations of the conformal group \cite{gross1964}, which includes the Poincar\'e group. These requirements led us to modify the usual definitions of plane waves and multipolar fields. Importantly, we show that the irregular incoming and outgoing multipolar fields transform under Lorentz boost as the regular multipolar fields, a new result that is needed for building the polychromatic T-matrix formalism. 

In the rest of the article we advance towards the definition of the polychromatic T-matrix and S-matrix in the following way. First, we define the basis states that are relevant in the T-matrix and S-matrix settings. Namely, regular multipolar fields for representing incident fields and irregular multipolar fields for representing incoming, outgoing, and scattered fields. To such end, we connect in Sec.~\ref{sec:connection} the traditional formalism of electric and magnetic fields $\left\{\bm E(\bm r,t),\bm B(\bm r,t)\right\}$ with the formalism based on abstract kets in a Hilbert space $\ket{f}$, and then define the plane wave states so that they transform according to the massless unitary representations of the Poincar\'e group with well defined helicity(handedness) $\lambda=+1$, or $\lambda=-1$. The regular multipolar fields are then defined from the plane waves. Both plane waves and multipoles defined in this way feature an extra factor of $k$ with respect to the usual definitions. Such factor ensures that both the fields and the expansion coefficients multiplying them transform unitarily under Lorentz boosts. For the case of incoming and outgoing multipolar fields, we are also led to multiply the usual definitions by a factor of $1/2$ by the notable properties of polychromatic irregular fields: namely that they vanish identically either before or after the light-matter interaction period, and are correspondingly equal to regular fields at certain time regions. We provide all the transformation properties for states and coefficients under the isometries of the Minkowski space-time: Poincar\'e transformations, parity, and time-reversal. The polychromatic T-matrix is defined in Sec.~\ref{sec:Tpol} and the polychromatic S-matrix is defined in Sec.~\ref{sec:Spol}. In Sec.~\ref{sec:diagonal} we consider the special case where the polychromatic T-matrix is diagonal in frequency, and show how to build it using monochromatic T-matrices that are computed with the usual conventions. As an exemplary application, the formalism is applied in Sec.~\ref{sec:Transfer} to the computation of the energy and linear momentum transferred from a pulse of light onto a silicon sphere.

All the numerical results contained in the paper can be reproduced with the code provided by request, together with the {\large \texttt{treams}} Python package \cite{beutel2021,Beutel2023}, which is publicly available at \url{https://github.com/tfp-photonics/treams}.

\section{Group theory-guided representations of electromagnetic fields \label{sec:connection}}
Maxwell equations in frequency domain for the electric $\tilde {\bm E}(\bm r, k)$  and magnetic fields $\tilde {\bm B}(\bm r, k)$ in vacuum and in SI units are
\eq{
	\bm \nabla \times \tilde {\bm E}(\bm r, k) &=i ck \tilde {\bm B}(\bm r, k),  & \bm \nabla \times \tilde {\bm B}(\bm r, k) &= -\frac{ i k}{c^2} \tilde {\bm E}(\bm r, k), \\
	\bm \nabla \cdot \tilde {\bm E}(\bm r, k) &= 0, & \bm \nabla \cdot \tilde {\bm B}(\bm r, k) &= 0,
}
where $c = 1/\sqrt{\epsilon_0 \mu_0}$ is the speed of light in vacuum and for convenience we describe frequency via the absolute value of the wavevector $ k =\sqrt{\bm k \cdot \bm k}=|\bm k|= \omega/c$. Since the magnetic field is completely determined by the electric field, we will focus on the latter for describing the electromagnetic field. 

It is convenient to start from the complex-valued electric field in the space-time domain, defined by setting the components of negative frequency in the Fourier decomposition of the field to zero:
\eq{
\bm E(\bm r, t) = \frac{1}{\sqrt{2\pi}}\int_0^{\infty} dk \, e^{-i kc t}\, \tilde{\bm E}(\bm r, k), \label{eq:Fcomplex}
}
while the real-valued electric field can be restored via
\eq{
	{\bm{\mathcal E}}( \bm r, t) = \bm E(\bm r, t)  + \bm E^*(\bm r, t) = 2 \Re[\bm E(\bm r, t)].
}
\subsection{Plane waves $|\mathbf{k} \lambda \rangle$}
One can decompose the electric field in plane waves of right-handed circular polarization, helicity $\lambda=  -1$, and of left-handed circular polarization, helicity $\lambda = 1$, using polarization vectors as defined in \cite{varshalovich1988} (Sec. 1.1.4)
\eq{
	\bm e_\lambda(\hat{\bm k}) &:=
	- \frac{1}{\sqrt{2}} (\lambda \bm e_\theta(\hat{\bm k})  + i\bm e_\phi(\hat{\bm k}) )\\
	 &=  \frac{1}{\sqrt{2}} 
	\begin{pmatrix} 
	- \lambda \cos\phi\cos\theta +  i \sin\phi \\ -\lambda \sin\phi\cos\theta - i\cos\phi \\ \lambda \sin\theta 
	\end{pmatrix}, \label{eq:ebasis}
}
where $\bm e_\theta$, $\bm e_\phi$ are spherical basis vectors, $\hat{\bm k}$ is the unit vector along the direction of the wave vector, with $\theta=\arccos\left(k_z/\abs{\bm k} \right)$ and $\phi=\arctantwo\left(k_y,k_x\right)$ being its polar and azimuthal angles.

The vectors $\bm e_\lambda(\hat{\bm k})$, $\lambda=\pm 1$ together with $\bm e_0(\hat{\bm k}) := \hat{\bm k}$ build a local orthonormal basis at $\bm k$. They are eigenvectors of the helicity operator $\Lambda = \frac{i\hbar\bm k \times }{k}$:
\eq{
	\frac{i \hbar \bm k \times }{k} \bm e_\lambda(\hat{\bm k}) = \lambda \hbar \bm e_\lambda(\hat{\bm k}) \ ,\qquad \text{ for } \lambda = -1, 0, 1.
} 
The $\lambda=\pm 1$ basis vectors are suitable for decomposition of transverse fields into parts of definite circular polarizations, since the $\lambda = 0$ fields have zero curl and do not occur in $k > 0$ Maxwell fields. To achieve the decomposition, first one performs the 3D Fourier transform of the complex electric field
\eq{
	\bm E(\bm r, t) &= \frac{1}{\sqrt{(2\pi)^3}}\int d^3 \bm k \, \bar{\bm E}(\bm k) \, e^{-i kc t} e^{i \bm k \cdot \bm r}, 
}
with absolute value of wave vector $k = \abs{\bm k} = \omega /c $, and then projects the polarization vectors of helicity $\lambda = \pm 1$ onto $\bar{\bm E}(\bm k)$, with dimensional constants chosen for future convenience:
\eq{
	f_\lambda(\bm k) =  \sqrt{2} \,\sqrt{\frac{ \epsilon_0}{c\hbar}}\, \bm e_\lambda(\hat{\bm k})^* \cdot \bar{\bm E}(\bm k),\label{eq:df}
}
so $\bar{\bm E}(\bm k) = \frac{1}{\sqrt{2}}\sqrt{\frac{c\hbar}{\epsilon_0}} \sum_{\lambda=\pm1} f_\lambda(\bm k) \bm e_\lambda(\hat{\bm k})$, which results in the decomposition 
\eq{
	\bm E(\bm r, t) =  \sqrt{\frac{c\hbar}{ \epsilon_0}} \frac{1}{\sqrt{2}}\frac{1}{\sqrt{(2\pi)^3}}\sum_{\lambda=\pm 1} \int \frac{d^3 \bm k}{k} \,  f_\lambda (\bm k) \,k\,\bm e_\lambda(\hat {\bm k})  \, e^{i (\bm k \cdot \bm r - c k t)}, \label{eq:decomp}
}
where the coefficients $f_\lambda(\bm k)$ obey \cite{moses1973} the transformation laws of a photon wave function.
The independent helicity $\lambda=\pm 1$ components (left- and right-handed polarization) of the electric field are the two Riemann-Silberstein vectors

$\bm F_\lambda(\bm r, t) = \big( \bm E(\bm r, t) + i\lambda c \bm B(\bm r, t) \big) / \sqrt{2}$:
\eq{
\bm F_\lambda(\bm r, t)  
 &= \frac{1}{\sqrt{(2\pi)^3}} \int d^3 \bm k \,\frac{ \bar{\bm E}(\bm k) + i \lambda c \bar{\bm B}(\bm k) }{\sqrt{2}}  e^{i (\bm k \cdot \bm r - c k t)} \label{eq:RS}\\
 &=   \frac{1}{\sqrt{2}}\frac{1}{\sqrt{(2\pi)^3}}  \int d^3 \bm k\,\Big( \bar{\bm E}(\bm k) + i \lambda   \frac{\bm k \times}{k}\bar{\bm E}(\bm k) \Big )  e^{i (\bm k \cdot \bm r - c k t)} \label{eq:RSilb} \\
 &=  \sqrt{\frac{c\hbar}{ \epsilon_0}} \frac{1}{\sqrt{2}}\frac{1}{\sqrt{(2\pi)^3}}  \int d^3 \bm k\,\frac{1}{\sqrt{2}}\Big( f_+(\bm k)\bm e_+(\hat{\bm k}) + f_-(\bm k)\bm e_-(\hat{\bm k}) \nonumber\\
 & \qquad+ \lambda  f_+(\bm k)\bm e_+(\hat{\bm k}) - \lambda f_-(\bm k)\bm e_-(\hat{\bm k}) \Big )  e^{i (\bm k \cdot \bm r - c k t)} \\
 & = \sqrt{\frac{c\hbar}{ \epsilon_0}}\frac{1}{\sqrt{(2\pi)^3}} \int\frac{ d^3 \bm k}{k}\, f_\lambda (\bm k)\bm e_\lambda (\hat{\bm k}) \, k \,  e^{i (\bm k \cdot \bm r - c k t)}, \label{eq:RSaux}
}
where we used the Maxwell equation in the wave vector space $ \bm k \times  \bar{\bm E}(\bm k)= ck\bar{\bm B}(\bm k)$ in the second line of \Eq{eq:RSaux}. Since we work with complex-valued electric and magnetic fields, the two Riemann-Silberstein vectors are independent and only together provide the complete description of the electromagnetic field: $\bm F_-$ for waves of the right-handed circular polarization and $\bm F_+$ for the left-handed.

We define the electromagnetic plane wave as  
\begin{equation}
	\boxed{
		\begin{aligned}
\ket{\bm k \,\lambda}&\equiv \bm Q_\lambda(\bm k,\bm r, t)\\ 
&=  \sqrt{\frac{c\hbar}{ \epsilon_0}}\, \frac{1}{\sqrt{2}} \frac{1}{\sqrt{(2\pi)^3}}\, k \, \bm e_\lambda(\hat{\bm k}) e^{- i kc t } e^{i \bm k \cdot \bm r} \label{eq:PW} 
		\end{aligned}
	}
\end{equation}

so the decomposition of the electromagnetic field is written as
\eq{
 \bm E(\bm r, t)&=\sum_{\lambda=\pm1 }\int \frac{d^3 \bm k}{k}\, f_\lambda (\bm k) \,\bm Q_\lambda(\bm k, \bm r, t). \label{eq:pw-measure}
}
 
The factor $k$ in the definition of the plane wave in Eq.~(\ref{eq:PW}) is important. We see that it appears because in \Eq{eq:decomp} we have changed from the integration measure $d^3 \bm k$ of the 3D Fourier transform, to the integration measure $\frac{d^3 \bm k}{k}$, which is the Lorentz invariant integration measure in the light cone \cite[Eq.~(2.5.15)]{Weinberg1995}\cite[Sec.~10.4.6]{tung1985}. This change then introduces a factor of $k$ in the definition of the plane wave in \Eq{eq:PW}, and, as we show in Sec.~\ref{sec:lbz}, it is precisely this factor of $k$ that makes $\bm Q_\lambda(\bm k,\bm r, t)$ and $f_\lambda (\bm k)$ transform as massless unitary irreducible representations of the \Poincare group with helicity $\lambda=\pm1$, which are the transformation properties that the photon wavefunction must have, according to Wigner's classification \cite{wigner1939}. The transformation rules are \cite[Eqs. (10.4-23), (10.4-24)]{tung1985}:
\begin{alignat}{2}
	T(a^\mu)  \ket{\bm k \,  \lambda} &= \ket{\bm k \,  \lambda} e^{-i a^\mu k_\mu } \label{eq:wkt1}\\
	R (\alpha, \beta, \gamma) \ket{\bm k \,  \lambda} &= \ket{\tilde{\bm k} \,  \lambda} e^{-i \lambda \psi },  \qquad &&\tilde{\bm k} = R (\alpha, \beta, \gamma) \bm k  \label{eq:wkt2}\\
	L_z(\xi) \ket{\bm k \,  \lambda} &= \ket{\bm k'\,\lambda},  \qquad &&\bm k' \; = L_z(\xi) \bm k \label{eq:wkt3}
\end{alignat}
where $T(a^\mu)$ is a 4D translation in Minkowski space (we use the convention of metric signature $(- + + +)$), so $a^\mu k_\mu = - a^0 \abs{\bm k} +  \bm a\cdot \bm k$. $R(\alpha, \beta, \gamma)=R_z(\alpha)R_y(\beta)R_z(\gamma)$ is a rotation operator with corresponding Euler angles, $\psi$ satisfies $R(0,0,\psi) =  R(\tilde\phi, \tilde\theta, 0)^{-1} R(\alpha, \beta, \gamma) R(\phi, \theta, 0)$ ($\tilde\phi, \tilde\theta$ are spherical angles of the rotated wave vector $\tilde{\bm k} = R (\alpha, \beta, \gamma) \bm k$), and $L_z(\xi)$ is a Lorentz boost (App.\ref{app:boosts}) along the z-axis with rapidity $\xi$. Since boosting along an arbitrary direction can be decomposed into a composition of rotations and a boost in the z-direction with Eq.~(\ref{eq:bdecomp}), we will only treat explicitly the boost along the z-direction.

Plane waves in this representation transform under parity and time reversal\footnote{
The time reversal is represented anti-unitarily by an operator $I_t$ satisfying $\braket{I_t \phi | I_t \psi} = \braket{\phi | \psi}^* = \braket{\psi | \phi}$.  } as
\eq{
	I_s \ket{ \bm k \lambda} &= \ket{ -\bm k \, -\lambda} \label{eq:qparity} \\
	I_t \ket{ \bm k \lambda} &= \ket{ -\bm k \, \lambda}.\label{eq:qtime}
}

The measure $\int \frac{d^3 \bm k}{k}$ is invariant under the action of the \Poincare group, which allows one to formulate the transformation of the field by transforming the coefficients $f_\lambda(\bm k)$ in a way similar to the basis vectors \cite[Secs.~7.6 and 10.5.1]{tung1985}. For example, for a boost in the z-direction Eq.~(\ref{eq:wkt3}) the invariance of the measure means that $\int \frac{d^3 \bm k}{|\bm k|}=\int \frac{d^3 L^{-1}_z(\xi)\bm k}{|L^{-1}_z(\xi)\bm k|}$, which, together with the linearity of the boost operator and \Eq{eq:wkt3} results in:
\eq{
	L_z(\xi) \ket{f} &= \sum_{\lambda=\pm1} \int \frac{d^3 \bm k}{k} \, f_\lambda (\bm k) \, \ket{ L_z(\xi) \bm k  \,\lambda} \nonumber \\
	&= \sum_{\lambda=\pm1} \int \frac{d^3 L^{-1}_z(\xi) \bm k}{\abs{L^{-1}_z(\xi) \bm k}} \, f_\lambda ( L^{-1}_z(\xi)\bm k) \, \ket{ \bm k  \lambda}\nonumber \\
	&= \sum_{\lambda=\pm1} \int \frac{d^3 \bm k}{k} \, f_\lambda ( L^{-1}_z(\xi)\bm k) \, \ket{ \bm k  \lambda}.
}
The rules for transforming the coefficients $f_\lambda(\bm k)$ are
\begin{alignat}{2}
	T(a^\mu)  f_\lambda(\bm k) &= f_\lambda(\bm k) e^{-i a^\mu k_\mu }, \label{eq:fwkt1}\\
	R (\alpha, \beta, \gamma) f_\lambda(\bm k) &= f_\lambda(\tilde{\bm {k}}) e^{-i \lambda \psi } ,  \qquad &&\tilde{\bm k} = R^{-1} (\alpha, \beta, \gamma) \bm k  \label{eq:fwkt2}\\
	L_z(\xi) f_\lambda(\bm k) &= f_\lambda(\bm k'),  \qquad &&\bm{k}' \; = L^{-1}_z(\xi) \bm k \label{eq:fwkt3}
\end{alignat}
with $\psi$ found this time from $R(0,0,\psi) =  R(\phi, \theta, 0)^{-1} R(\alpha, \beta, \gamma) R(\tilde \phi, \tilde \theta, 0)$, where $\tilde\phi, \tilde\theta$ belong to the wave vector $\tilde{\bm k} = R^{-1} (\alpha, \beta, \gamma) \bm k$. The transformations under parity and time reversal are
\eq{
	I_s f_\lambda(\bm k) &= f_{-\lambda}(-\bm k)\\
	I_t f_\lambda(\bm k) &= f^*_\lambda(-\bm k).
}

Transformation properties of the defined plane waves and coefficients in the decomposition Eq.~(\ref{eq:pw-measure}) justify the way of writing the electric field as a ket
\eq{
	\ket{f} = \sum_{\lambda=\pm1} \int \frac{d^3 \bm k}{k} \, f_\lambda (\bm k) \, \ket{\bm k \lambda}. \label{eq:state}
}
The coefficients $f_\lambda (\bm k)$ belong to a Hilbert space with the scalar product 
\eq{
	\braket{f|g} = \sum_{\lambda=\pm1} \int \frac{d^3 \bm k}{k} \, f^*_\lambda (\bm k) g_\lambda (\bm k). \label{eq:scalarPW}
}
Such Hilbert space is isomorphic to the Hilbert space of solutions of Maxwell equations.

Thanks to the extra dimensional factors in \Eq{eq:df}, the $f_\lambda(\bm k)$ have the units inverse to the wavevector, that is, meters: $[f_\lambda(\bm k)]=\SI{}{\meter}$. The scalar product Eq.~(\ref{eq:scalarPW}) is then dimensionless, which is consistent with the physical interpretation of $\braket{f|f}$ as the number of photons \cite{zeldovich1965} contained in the field described by $\ket{f}$. The integral in \cite[Eq.~(1)]{zeldovich1965} is a different representation of the same scalar product, as can be seen by comparing Eq.~(3) and Eq.~(6) in \cite{gross1964}.

The scalar product also allows to quantify fundamental properties that are carried by the field, for example, energy, linear momentum, and angular momentum - using the expectation values $\braket{f | \Gamma| f}$ with $\Gamma$ being the generator of the corresponding symmetry transformation: time translation for energy, spatial translation for linear momentum and rotation for angular momentum.

It is also known\cite{gross1964} that the scalar product in Eq.~(\ref{eq:scalarPW}) is invariant under the conformal group, which is the largest group of invariance of Maxwell equations \cite{Bateman1909}. It contains the Poincar\'e group, and additionally special conformal transformations and dilations. 

Appendix~\ref{sec:appA} contains a brief discussion about the representation of the vector potential and its transformation properties. In particular, the plane waves for decomposing the vector potential {\em do not} feature the extra factor of $k$.

\subsubsection{Lorentz boosts $L_z(\xi)\ket{\mathbf{k} \lambda}$ \label{sec:lbz}}

We provide here the explicit derivation of the transformation of $\bm Q_\lambda(\bm k, \bm r, t)$ upon an \textit{active} boost of in $z$-direction with rapidity $\xi$. The rest of the transformation properties in Eqs.~(\ref{eq:wkt1},\ref{eq:wkt2},\ref{eq:qparity},\ref{eq:qtime}) are derived in App.~\ref{sec:Qproperties}.

The transformation of general electromagnetic fields in \cite[Eq.~(11.149)]{Jackson1998} can be used to derive the action of the boost (see also Eq.~(\ref{eq:boost})):
\eq{
	\bm Q_\lambda(\bm k, \bm r, t) &= \gamma \bm Q_\lambda(\bm k, \tilde{\bm r}, \tilde t) + \frac{i\lambda \gamma}{c} \bm v \times \bm Q_\lambda(\bm k, \tilde{\bm r}, \tilde t) - \frac{\gamma^2 \bm v}{(\gamma+1)c^2} \bm v \cdot \bm Q_\lambda(\bm k, \tilde{\bm r}, \tilde t)\nonumber  \\
	&=\Big (\gamma \mathds{1} + \frac{i\lambda \gamma}{c} \bm v \times - \frac{\gamma^2 \bm v}{(\gamma+1)c^2} \bm v \cdot \Big) \bm Q_\lambda(\bm k, \tilde{\bm r}, \tilde t),  \label{eq:pwb0}
}
with $\bm v = v \bm e_z = c \tanh(\xi) \bm e_z$, $\gamma = (1-v^2/c^2)^{-1/2} = \cosh(\xi)$, $\gamma v = c \sinh(\xi)$ and inversely transformed space-time coordinates
\eq{
	\begin{pmatrix}  
		c\tilde t \\ \tilde{\bm r} 
	\end{pmatrix}
	=
L^{-1}_z(\xi) 
	\begin{pmatrix}  
		c t \\ \bm r 
	\end{pmatrix}
= 
\begin{pmatrix}
\cosh(\xi) & 0 & 0 & -\sinh(\xi) \\
0 & 1 & 0 & 0 \\
0 & 0 & 1 & 0 \\
-\sinh(\xi) & 0 & 0 & \cosh(\xi) \\
\end{pmatrix}
	\begin{pmatrix}  
		c t \\ x \\ y \\ z
	\end{pmatrix}. \label{eq:pwb1}
}
It is important to distinguish between the passive and active versions of the Lorentz boost. In the passive version, where the reference frame is boosted instead of the field, Eqs.~(\ref{eq:pwb0}-\ref{eq:pwb1}) incorporate $-\bm v$ in place of $\bm v$ (and, equivalently, $-\xi$ instead of $\xi)$.

Since the helicity basis vectors $\bm e_\sigma(\hat{\bm k})$ can be obtained as $\bm e_\sigma(\hat{\bm k}) = R(\phi,\theta,0) \bm e_\sigma(\hat{\bm z})$, and they transform under rotations as 
\eq{
	 R(\alpha, \beta, \gamma) \, \bm e_\lambda( \hat{\bm k}) = \sum_{\sigma = \pm 1, 0} D^1_{\sigma \lambda}(\alpha,\beta,\gamma)\, \bm e_\lambda(\hat{\bm k})
}
where $D^j_{mn}(\alpha,\beta,\gamma) = e^{-im\alpha} d^j_{mn}(\beta)e^{-in\gamma}$ are Wigner matrices and $d^j_{mn}(\beta)$ are small Wigner matrices, as defined in \cite{tung1985}, Sec. 7.3. Then:
\eq{
	&\Big (\gamma \mathds{1} + \frac{i\lambda \gamma}{c} \bm v \times - \frac{\gamma^2 \bm v}{(\gamma+1)c^2} \bm v \cdot \Big) \,  k\, \bm e_\lambda( \hat{\bm k}) \nonumber \\
	&=\Big (\cosh(\xi) \mathds{1} + i\lambda \sinh(\xi) \bm e_z \times - \frac{\sinh^2(\xi) \bm e_z}{\cosh(\xi)+1} \bm e_z \cdot \Big) \sum_{\sigma=-1,0,1}\! D^1_{\sigma\lambda}(\phi,\theta,0)\,  k \, \bm e_\sigma( \hat{\bm z}) \nonumber \\
	&= \sum_{\sigma=-1,0,1} D^1_{\sigma\lambda}(\phi,\theta,0)\Big (\cosh(\xi) + \lambda\sigma \sinh(\xi) - \frac{\sinh^2(\xi) \delta_{0\sigma}}{\cosh(\xi)+1} \Big) \,  k\, \bm e_\sigma( \hat{\bm z}) \label{eq:pw_b1} \\
	&= \sum_{\sigma=-1,0,1} e^{-i \phi\sigma } \, d^1_{\sigma\lambda}(\theta)\Big ( \sigma\lambda \sinh(\xi)+  \frac{\cosh(\xi) + \cosh^2(\xi)- \sinh^2(\xi) \delta_{0\sigma}}{\cosh(\xi)+1} \Big) \,  k\, \bm e_\sigma( \hat{\bm z}) \nonumber \\	
	&= \sum_{\sigma=-1,0,1} e^{-i \phi\sigma } \, d^1_{\sigma\lambda }(\tilde \theta)\, \tilde k \, \bm e_\sigma( \hat{\bm z}) \label{eq:pw_b2}\\
	&= \tilde k\, \bm  e_\lambda(\hat{\tilde{\bm k}}).
	}
For Eq.~(\ref{eq:pw_b1}) we have used $\bm e_z \times \bm e_\sigma(\hat{\bm z}) = -i \sigma \bm e_\sigma(\hat{\bm z})$ and $\bm e_z  \big(\bm e_z \cdot \bm e_\sigma(\hat{\bm z}) \big) = \delta_{0\sigma} \bm e_\sigma(\hat{\bm z})$ for $\sigma = -1,0,1$. Eq.~(\ref{eq:pw_b2}) follows from the transformation rules for the wave vector $\bm k$ (App.\ref{app:boosts}):
\eq{
	\tilde k \, d^1_{0\lambda}(\tilde \theta) = \frac{\lambda \tilde k}{\sqrt{2}} \sin(\tilde\theta)  = \frac{\lambda k}{\sqrt{2}} \sin(\theta)  
	= k \, d^1_{0\lambda}(\theta)
}
for $\lambda=\pm 1$, $\sigma= 0$ and	
\eq{
	\tilde k\, d^1_{\sigma\lambda}(\tilde \theta) &= \frac{\tilde k}{2}\, \big(1 + \sigma\lambda \cos(\tilde\theta)\big) \\
	 &= \frac{ k \big(\cosh(\xi) + \cos(\theta) \sinh(\xi)\big) }{2}\, \Big(1 + \sigma\lambda  \frac{\cos(\theta)\cosh(\xi) + \sinh(\xi)} {\cosh(\xi)  + \cos(\theta) \sinh(\xi) } \Big) \nonumber\\
	  &= \frac{ k }{2}\, \big(\cosh (\xi)  + \cos(\theta) \sinh(\xi) + \sigma\lambda  (\cos(\theta)\cosh(\xi) + \sinh(\xi)) \big)\nonumber \\
	 &= \frac{ k }{2}\, \big(1 +  \sigma\lambda \cos(\theta)\big)\big(\cosh(\xi) + \sigma\lambda \sinh(\xi) \big)\nonumber \\
	 &= k\, d^1_{\sigma\lambda}(\theta) \big( 1 + \sigma\lambda \sinh(\xi) \big)
}
for $\lambda = \pm 1$, $\sigma=\pm 1$.\\
Together with the fact that for any Lorentz boost $L$ 
\eq{
 	e^{-i k^\mu(L^{-1} x)_\mu} = e^{-i (L k)^\mu x_\mu},
}
this implies Eq.~(\ref{eq:wkt3}):
\eq{
	\boxed{\Big (\gamma \mathds{1} + \frac{i\lambda \gamma v}{c} \bm e_z \times - \frac{\gamma^2 v^2 \bm e_z}{(\gamma+1)c^2} \bm e_z \cdot \Big)\bm Q_\lambda(\bm k, \tilde{\bm r},\tilde t) = \bm Q_\lambda(\tilde{\bm k}, \bm r, t).}
}

We emphasize that without the $k$-factor in the definition Eq.~(\ref{eq:PW}), the plane wave would not transform according to the unitary representation Eq.~(\ref{eq:wkt3}).

\subsection{Angular momentum basis for regular fields $|k j m \lambda\rangle$}
The multipolar fields, also known as vector spherical harmonics or angular momentum fields, play a crucial role in the T-matrix formalism: they constitute the basis with respect to which the fields are expanded. 

The angular momentum basis can be defined with respect to the plane wave basis as\cite[Sec. 8.4.1]{tung1985}:
\eq{
	\ket{k j m \lambda} &= \sqrt{\frac{2j+1}{4\pi}} \int^{2\pi}_0 \!d\phi \, \int_{-1}^{1} \! d(\cos \theta) \, D^j_{m\lambda}(\phi,\theta,0)^* \ket{\bm k \lambda}\label{eq:AM}\\
	\ket{\bm k \lambda} &=  \sum_{j=1}^{\infty} \sum_{m=-j}^j \sqrt{\frac{2j+1}{4\pi}} D^j_{m\lambda}(\phi,\theta,0) \ket{k j m \lambda}
	\label{eq:PW2}
}
with the corresponding connection between coefficients in the angular momentum and the plane wave basis
\eq{
	f_{j m \lambda}(k) &= \sqrt{\frac{2j+1}{4\pi}} \int^{2\pi}_0 \!d\phi \, \int_{-1}^{1} \! d(\cos \theta) \, D^j_{m\lambda}(\phi,\theta,0)\, f_\lambda(\bm k)\\
	f_\lambda(\bm k) &=  \sum_{j=1}^{\infty} \sum_{m=-j}^j \sqrt{\frac{2j+1}{4\pi}} D^j_{m\lambda}(\phi,\theta,0)^* \, f_{j m \lambda}(k). \label{eq:lasteq}
}

The indices in $\ket{k j m \lambda}$ correspond to eigenvalues of Hermitian operators of energy $H$, total angular momentum $J^2=J_x^2+J_y^2+J_z^2$, angular momentum along $z$-axis $J_z$, and helicity $\Lambda$:
\begin{align*}
	H \ket{k j m \lambda}&=\hbar ck\ket{k j m \lambda} \\
	J^2\ket{k j m \lambda}&=\hbar^2 j(j+1)\ket{k j m \lambda}, \quad j=1,2,\ldots\\
	J_z\ket{k j m \lambda}&=\hbar m\ket{k j m \lambda}, \qquad \qquad \!\! m=-j,-j+1,\ldots,j\\
	\Lambda\ket{k j m \lambda}&=\hbar\lambda\ket{k j m \lambda}, \qquad \qquad \,\, \lambda= \pm  1.
\end{align*}

In particular, $j=1$ corresponds to the dipolar fields, $j=2$ corresponds to the quadrupolar fields, and so on. We remark for completeness that the plane waves are eigenstates of helicity $\Lambda\ket{\bm k \lambda}=\hbar\lambda\ket{\bm k \lambda}$, and of the three translation operators, and hence of their Hermitian generators, the linear momentum operators $P_{x,y,z}\ket{\bm k \lambda}=\hbar k_{x,y,z}\ket{\bm k \lambda}$.

A state can be represented in the angular momentum basis by using Eqs.(\ref{eq:AM}-\ref{eq:lasteq}) in  Eq.~(\ref{eq:state}), which results in
\eq{
	\ket{f} &=  \int_0^\infty dk \, k \, \sum_{\lambda=\pm 1} \sum_{j=1}^{\infty} \sum_{m=-j}^j \, f_{j m \lambda}(k) \, \ket{k j m \lambda}\label{eq:AMref}
}
and the scalar product in \Eq{eq:scalarPW} can correspondingly be written in the angular momentum basis as
\eq{
	\braket{g | f} &= \sum_{\lambda=\pm 1} \int_0^{\infty} dk\, k \sum_{j=1}^{\infty} \sum_{m=-j}^j \, g_{j m \lambda}(k)^* f_{jm\lambda}(k). \label{eq:scalarAM}
}
We note that coefficients $f_{jm\lambda}(k)$ have the units of meters just as $f_\lambda(\bm k)$. 

This allows one to obtain explicit $(\bm r,t)$-dependent expressions of the angular momentum basis $\ket{kjm\lambda}$ for the regular electromagnetic field as follows:
\eq{
	&\bm R_{jm\lambda}(k,\bm r, t) =\sqrt{\frac{2j+1}{4\pi}} \int^{2\pi}_0 \!d\phi \, \int_{-1}^{1} \! d(\cos \theta) \, D^j_{m\lambda}(\phi,\theta,0)^* \bm Q_\lambda(\bm k,\bm r, t)\nonumber \\
	&= \sqrt{\frac{c\hbar}{\epsilon_0}} \frac{k \, e^{-i kc t}}{\sqrt{2}\sqrt{(2\pi)^3}} \,\sqrt{\frac{2j+1}{4\pi}} \int^{2\pi}_0 \!d\phi \, \int_{-1}^{1} \! d(\cos \theta) \, D^j_{m\lambda}(\phi,\theta,0)^* \bm e_\lambda(\hat{\bm k}) e^{i \bm k \cdot \bm r},
}
where the integration proceeds over the polar and azimuthal angles ($\theta$, $\phi$) of the wave vector. The integration results in
\eq{
	\bm R_{j m\lambda}(k, \bm r, t) 	&= \sqrt{\frac{c\hbar}{\epsilon_0}} \frac{k \, e^{-i kc t}}{\sqrt{\pi}\sqrt{2j+1}}  \sum_{L=j-1}^{j+1} \sqrt{2L + 1}\, i^{L} j_{L}(k r) \, C^{j\lambda}_{L0,1\lambda} \bm Y^L_{j m}(\hat{\bm r}) 
	 \label{eq:defR} \\
	&\equiv \ket{k j m \lambda} \nonumber
}
with Clebsch-Gordan coefficients $C^{j_3 m_3}_{j_1 m_1 j_2 m_2}$ and vector spherical harmonics\cite[Sec. 7.3.1]{varshalovich1988}
\eq{
	\bm Y^L_{j m}(\hat{\bm r}) &= \sqrt{\frac{2L+1}{4\pi}}\sum_{\sigma=\pm1,0}\bm e_\sigma(\hat{\bm z})  D^L_{m-\sigma, 0}(\phi, \theta, 0)^*\, C^{jm}_{L m-\sigma,1\sigma}.
}
The decomposition of the regular electromagnetic field $\bm E(\bm r, t)$ then reads, as suggested by \Eq{eq:AMref}
\eq{
	\bm E(\bm r, t) &= \int_0^{\infty} dk \, k \sum_{\lambda=\pm 1} \sum_{j=1}^{\infty}  \sum_{m=-j}^j \, f_{j m \lambda}(k) \bm R_{jm\lambda}(k, \bm r, t).  
}

The connection between $\bm R_{jm\lambda}(k, \bm r, t)$ and the usual regular electric and magnetic multipoles 
\eq{
	\bm N_{jm}(k r, \hat{\bm r}) &= i  j_{j-1} (k r) \sqrt{\frac{j+1}{2j+1}} \bm Y^{j-1}_{j m}(\hat{\bm r})  -i  j_{j+1} (k r) \sqrt{\frac{j}{2j+1}} \bm Y^{j+1}_{j m}(\hat{\bm r}) \label{eq:regmultN}\\
	\bm M_{jm}(k r, \hat{\bm r})  &= j_j (k r) \bm Y^j_{j m}(\hat{\bm r}) \label{eq:regmultM} 
}
can be found using the expression for Clebsch-Gordan coefficients
\eq{
C^{j\lambda}_{L0,1\lambda} &=
 \begin{cases} 
 \sqrt{\frac{j}{2(2j+3)}}, & \text{if } L=j+1 \\ 
 -\frac{\lambda}{ \sqrt{2}}, & \text{if } L=j \\
 \sqrt{\frac{(j+1)}{2(2j-1)}}, & \text{if } L=j-1
 \end{cases}	
}
for $\lambda=\pm 1$. The relation is then

\begin{equation}
   \boxed{
   \begin{aligned}
	  \ket{k j m \lambda}&\equiv \bm R_{j m\lambda}(k, \bm r, t)\\
	&= -  \sqrt{\frac{c\hbar}{\epsilon_0}} \frac{1}{\sqrt{2\pi}} \, k \, i^j  \Big(  e^{-i kc t}\, \bm N_{jm}(k r, \hat{\bm r}) + \lambda \,e^{-i kc t} \, \bm M_{jm}(k r, \hat{\bm r} ) \Big) \label{eq:relation}
   \end{aligned}
   }
\end{equation}

with corresponding inverse relations
\eq{
e^{-ikc t}\, \bm N_{jm}(k r, \hat{ \bm{r}}) &= -\sqrt{\frac{\epsilon_0}{c\hbar}} \frac{1}{2} \Big( \bm R_{jm+}(k, \bm r, t) + \bm R_{jm-}(k, \bm r, t) \Big) \frac{(-i)^{j}\sqrt{2\pi}}{k} \\
e^{-ikc t}\, \bm M_{jm}(k r, \hat{ \bm{r}}) &= -\sqrt{\frac{\epsilon_0}{c\hbar}} \frac{1}{2} \Big( \bm R_{jm+}(k, \bm r, t) - \bm R_{jm-}(k, \bm r, t) \Big) \frac{(-i)^{j}\sqrt{2\pi}}{k}.
}

We note that the factor of $k$ in the plane wave definition of Eq.~(\ref{eq:PW}) leads to the extra factor of $k$ in the angular momentum basis in Eq.~(\ref{eq:relation}) compared to the usual multipolar basis $\bm M$ and $\bm N$. Such difference ensures that $\bm R_{j m\lambda}(k, \bm r, t)$ have the same transformation laws as $\ket{kjm\lambda}$ under the action of the Poincar\'e group (see Sec.(\ref{sec:AMlist})). The $\ket{kjm\lambda}$ transform unitarily in particular under Lorentz boosts because the $\ket{kjm\lambda}$ are unitarily connected to $\ket{\bm k \lambda}$ by Eq.~(\ref{eq:AM}), and we have already shown that the $\ket{\bm k \lambda}$ transform unitarily under Lorentz boosts.

Computing the T-matrix of an object moving with constant speed from the T-matrix of the object at rest requires the Lorentz boost of the T-matrix at rest. The T-matrix connects regular incident fields with irregular outgoing fields, and one cannot a priori assume that the two kinds of fields transform identically under boosts. Yet, that is indeed the case, as we proof in Sec.~\ref{sec:irrboost}. Therefore, knowledge of the boost matrix in the regular angular momentum basis, which we derive in Sec.~\ref{sec:boost-matrix}, is sufficient for boosting T-matrices. The transformation properties of the $\ket{kjm\lambda}$ under other transformations can be found in Sec.~\ref{sec:AMlist}. 

\subsubsection{Matrix element of Lorentz boosts $\braket{k_1 j_1 m_1 \lambda_1 |L_z(\xi)|k_2 j_2 m_2 \lambda_2}$}\label{sec:boost-matrix}
We are after the matrix element of a Lorentz boost along the z-direction for massless unitary irreducible representations of the Poincar\'e group in angular momentum basis from the boost law of the plane wave  Eq.~(\ref{eq:wkt3}) and the connection between the plane-wave and angular momentum basis Eqs.(\ref{eq:AM})-(\ref{eq:PW2}).
\eq{
	&L_z(\xi) \ket{k_2 j_2 m_2 \lambda_2} = \nonumber \\
	&= L_z(\xi) \int_0^\infty dk \, k \, \frac{1}{k_2} \delta(k -k_2) \ket{k j_2 m_2 \lambda_2} \nonumber \\
	&= L_z(\xi) \int_0^\infty d k \, k \, \frac{1}{k_2} \delta(k-k_2)  \int^{2\pi}_0 \!d\phi \, \int_{-1}^{1} \! d(\cos \theta) \, \sqrt{\frac{2j_2+1}{4\pi}}\, D^{j_2}_{m_2 \lambda_2}(\phi, \theta, 0)^* \ket{\bm k \lambda_2} \nonumber \\
	&= \sqrt{\frac{2j_2+1}{4\pi}} \int \frac{d^3 \bm k}{k} \frac{1}{k_2} \delta(k-k_2)  D^{j_2}_{m_2 \lambda_2}(\phi, \theta, 0)^* \ket{\tilde{\bm k} \lambda_2}
}
with boosted wave vector according to  Eq.~(\ref{eq:wkt3})
\eq{
\tilde{\bm k} = L_z(\xi) \bm k = 
\begin{pmatrix}
	k^1 \\ k^2 \\ k \sinh(\xi) + k^3 \cosh(\xi)
\end{pmatrix}.
}
Positive rapidity $\xi$ corresponds to the active boost in the positive z-direction, and vice versa for the negative $\xi$. Using invariance of the measure
\eq{
	 \int \frac{d^3 \bm k}{k} \, f_\lambda(\bm k) \ket{L_z(\xi) \bm k, \lambda} =  \int \frac{d^3 \bm k}{k} \, f_\lambda(L^{-1}_z(\xi) \bm k) \ket{\bm k \lambda} 
}
one continues with
\eq{
&L_z(\xi) \ket{k_2 j_2 m_2 \lambda_2} = \sqrt{\frac{2j_2+1}{4\pi}} \int \frac{d^3 \bm k}{k} \frac{1}{k_2} \delta(k' -k_2)  D^{j_2}_{m_2 \lambda_2}(\phi', \theta', 0)^* \ket{ \bm k \lambda_2},
}
where the absolute value and the spherical angles of the inversely boosted wave vector are (App. \ref{app:boosts})
\eq{
	&\cos(\theta') = \frac{\cos(\theta) - \tanh(\xi)}{1 - \cos(\theta) \tanh(\xi)},\\
	&\phi' = \phi, \\
	&k' = k \big( \cosh(\xi) - \cos(\theta) \sinh(\xi) \big) .
}
Then
\eq{
&L_z(\xi) \ket{k_2 j_2 m_2 \lambda_2} \nonumber \\
&= \sqrt{\frac{2j_2+1}{4\pi}} \int \frac{d^3 \bm k}{k} \frac{1}{k_2} \delta(k' -k_2)  D^{j_2}_{m_2 \lambda_2}(\phi, \theta', 0)^* \ket{ \bm k \lambda_2} \nonumber \\
&=\sqrt{\frac{2j_2+1}{4\pi}} \int_0^\infty dk \, k \int^{2\pi}_0 \!d\phi \, \int_{-1}^{1} \! d(\cos \theta) \frac{1}{k_2} \delta(k' -k_2)  D^{j_2}_{m_2 \lambda_2}(\phi, \theta', 0)^* \ket{ \bm k \lambda_2 \nonumber} \\
&= \sqrt{\frac{2j_2+1}{4\pi}} \int_0^\infty dk \, k \int^{2\pi}_0 \!d\phi \, \int_{-1}^{1} \! d(\cos \theta) \frac{1}{k_2} \delta(k' -k_2)  D^{j_2}_{m_2 \lambda_2}(\phi, \theta', 0)^* \nonumber \\
&\quad\times\sum_{j=1}^\infty\sum_{m=-j}^j\sqrt{\frac{2j+1}{4\pi}}\,D^{j}_{m \lambda_2}(\phi, \theta, 0) \ket{k j m \lambda_2}\nonumber\\
&= 2\pi\,\sqrt{\frac{2j_2+1}{4\pi}}\int_0^\infty dk \, k\, \int_{-1}^{1} \! d(\cos \theta) \frac{1}{k_2} \delta(k' -k_2)  d^{j_2}_{m_2 \lambda_2}(\theta') \sum_{j=1}^\infty\sqrt{\frac{2j+1}{4\pi}} \,d^{j}_{m_2 \lambda_2}(\theta) \ket{k j m_2 \lambda_2},
}
Where in the last step we integrated over $\phi$. For compactness of the formulas we imply here and in the following that small Wigner functions $d^{j}_{m_2 \lambda_2}(\theta)$ are zero for $j < \abs{m_2}$ rather than undefined. Rewriting
\eq{
	\delta(k' - k_2) &= \delta(k \cosh(\xi) - k \cos(\theta) \sinh(\xi) - k_2) \nonumber \\
	&= \delta\Big( k \sinh(\xi) \big(\cos\theta - \frac{k\cosh(\xi) - k_2}{k \sinh{\xi}}\big)\Big) \nonumber \\
	&=\frac{1}{k \abs{\sinh(\xi)}} \, \delta\Big( \cos\theta - \frac{k \cosh \xi - k_2}{k \sinh \xi}\Big),
}
integrating over $\cos\theta$ and renaming variables $j\rightarrow j_1$, $k \rightarrow k_1$ gives
\eq{
& L_z(\xi) \ket{k_2 j_2 m_2 \lambda_2} = \nonumber \\
&= \frac{1}{2}\sqrt{2j_2+1}  \int_{k_2 e^{-\abs{\xi}}}^{k_2 e^{\abs{\xi}}} dk_1 \, k_1 \sum_{j_1=1}^\infty \sqrt{2j_1+1} \frac{1}{k_2k_1 \abs{\sinh(\xi)}}d^{j_1}_{m_2 \lambda_2}(\theta_1) d^{j_2}_{m_2 \lambda_2}(\theta_2) \ket{k_1 j_1 m_2 \lambda_2} \nonumber \\
&= \frac{1}{2}\sqrt{2j_2+1}  \int_{k_2 e^{-\xi}}^{k_2 e^{\xi}} dk_1 \, k_1 \sum_{j_1=1}^\infty \sqrt{2j_1+1} \frac{1}{k_2k_1 \sinh(\xi)}d^{j_1}_{m_2 \lambda_2}(\theta_1) d^{j_2}_{m_2 \lambda_2}(\theta_2) \ket{k_1 j_1 m_2 \lambda_2}
\label{eq:boostedAM}
}
with $\theta_1$ and $\theta_2$ defined via
\eq{
\cos \theta_1 = \frac{k_1 \cosh \xi - k_2}{k_1 \sinh \xi},\\
\cos \theta_2 = \frac{k_1 - k_2 \cosh \xi}{k_2 \sinh \xi}.
}
The Eq.~(\ref{eq:boostedAM}) is the closed form expression equivalent to Eq.~(5.15) in \cite{moses1965}. 

To write the matrix element of $L_z(\xi)$ in the angular momentum basis we account for the integration limits with a Heaviside function $\Theta$
\eq{
& L_z(\xi) \ket{k_2 j_2 m_2 \lambda_2} = \nonumber\\
&= \frac{1}{2}\sqrt{2j_2+1} \int_0^\infty dk_1 \, k_1 \sum_{j_1}\sqrt{2j_1+1} \frac{1}{k_2 k_1 \abs{\sinh(\xi)}}\,d^{j_1}_{m_2 \lambda_2}(\theta_1)\, d^{j_2}_{m_2 \lambda_2}(\theta_2)\, \Theta\big(\abs{\xi} - \abs{\ln(k_1/k_2)} \big) \, \ket{k_1 j_1 m_2 \lambda_2}
}
where
\eq{
\Theta(x) = 
\begin{cases} 1, & \text{if } x\geq 0 \\ 0, & \text{if } x < 0  \end{cases}.
}
The matrix element is then
\begin{equation}
   \boxed{
   \begin{aligned}
	   &\braket{k_1 j_1 m_1 \lambda_1 |L_z(\xi)|k_2 j_2 m_2 \lambda_2} =\\
	   &\frac{1}{2}\,\sqrt{2j_1+1}\,\sqrt{2j_2+1}\, \frac{\Theta\big(\abs{\xi} - \abs{\ln(k_1/k_2)} \big)}{k_1k_2 \abs{\sinh\xi}}\, d^{j_1}_{m_1 \lambda_1}(\theta_1) \, d^{j_2}_{m_2 \lambda_2}(\theta_2)  \delta_{m_1 m_2}\delta_{\lambda_1 \lambda_2}.\label{eq:matelboost}
   \end{aligned}
   }
\end{equation}

Applying to the integration over $k_1$ in Eq.~(\ref{eq:boostedAM}) a substitution
\eq{
	k_1 = k_2\cosh(\xi) + k_2 \cos(\theta_2)\sinh(\xi)
}
together with renaming of variables $\theta_2 \rightarrow \theta$, $k_2\rightarrow k$, $j_2\rightarrow j$, $k_1\rightarrow k'$, $j_1\rightarrow j'$  brings the transformed basis state to the form that we will also find useful:
\eq{
 L_z(\xi) \ket{k j m \lambda} = \frac{1}{2}\,\sqrt{2j+1}\sum_{j'=1}^{\infty} \,\sqrt{2j'+1} \int^{1}_{-1} 
 d(\cos \theta) \, d^j_{m\lambda}(\theta) \,  d^{j'}_{m\lambda}(\theta')     \ket{k' j' m \lambda},  \label{eq:useful}
}
where
\eq{
\cos(\theta') &= \frac{\cos(\theta) + \tanh(\xi)}{1 + \cos(\theta) \tanh(\xi)}\label{eq:thetaprim}
}
and
\eq{
k' = k\big(\cosh(\xi) + \cos(\theta) \sinh(\xi)\big). \label{eq:kprim}
}

The derived laws describe the transformation property of basis vector fields $\bm R_{jm\lambda}(k, \bm r, t)$ for all space-time points $(\bm r, t)$.
\subsubsection{Other transformations of $\ket{kjm\lambda}$}\label{sec:AMlist}
The complete table of transformation laws of the electric field under the isometries of the Minkowski space-time reads:
\eq{
&T_t(\tau) \ket{k j m \lambda} = \ket{k j m \lambda} e^{ikc\tau} \label{eq:AMtimetr}\\
&T_z(a) \ket{k j m \lambda} = \sum_{j'=1}^{\infty} \sqrt\frac{2j' + 1}{2j +1} \sum_{l=\abs{j-j'}}^{j+j'} \, (2l + 1) (-i)^l j_l(ak) C^{jm}_{j'm,l0}C^{j\lambda}_{j'\lambda,l0}\ket{k j' m \lambda}\label{eq:AMztr},\\
&R(\alpha, \beta, \gamma) \ket{k j m \lambda} =  \sum_{m'=-j}^j D^j_{m' m}(\alpha, \beta, \gamma) \ket{k j m' \lambda} \\
 &L_z(\xi) \ket{k j m \lambda} = \frac{1}{2}\,\sqrt{2j+1}\sum_{j'=1}^{\infty} \,\sqrt{2j'+1} \int^{1}_{-1} d(\cos \theta)  \, d^j_{m\lambda}(\theta) \,  d^{j'}_{m\lambda}(\theta')  \ket{k' j' m \lambda}
}
where $\theta'$ and $k'$ are given by Eqs.(\ref{eq:thetaprim}-\ref{eq:kprim}). $T_t(\tau)$ is the time translation by $\tau$,  $T_z(a)$ is the translation in the positive z-direction by $a$, and the translation in the general direction $\hat{\bm n}(\alpha,\beta)$ can be described by $T_{\hat{\bm n}}(\xi) = R(\alpha, \beta, 0) T_z(a) R^{-1}(\alpha,\beta,0)$ in the similar way as general Lorentz boosts.\\
The actions of parity and time reversal are given by (See App.~\ref{app:AMparity})
\eq{
	I_s \ket{k j m \lambda} &= \ket{k j m -\lambda}(-1)^j \\
	I_t \ket{k j m \lambda} &= -\ket{k j -m \lambda}(-1)^{j+m}.
}
The corresponding rules for transformations of coefficients are
\eq{
&T_t(\tau) f_{j m \lambda}(k) = f_{j m \lambda}(k) e^{ikc\tau} \\
&T_z(a) f_{j m \lambda}(k) = \sum_{j'=1}^{\infty} \sqrt\frac{2j + 1}{2j' +1} \sum_{l=\abs{j-j'}}^{j+j'} \, (2l + 1) (-i)^l j_l(ak) C^{j'm}_{jm,l0}C^{j'\lambda}_{j\lambda,l0} \, f_{j' m \lambda}(k),\\
&R(\alpha, \beta, \gamma) f_{j m \lambda}(k) =  \sum_{m'=-j}^j D^j_{m m'}(\alpha, \beta, \gamma) \,f_{j m' \lambda}(k), \\
&L_z(\xi) f_{j m \lambda}(k) = \frac{1}{2}\,\sqrt{2j+1}\sum_{j'=1}^{\infty} \,\sqrt{2j'+1} \int^{1}_{-1} d(\cos \theta) \, d^j_{m\lambda}(\theta) \,  d^{j'}_{m\lambda}(\theta') f_{j' m \lambda}(k')
}
with $\theta'$ and $k'$ given by 
\eq{
\cos(\theta') &= \frac{\cos(\theta) - \tanh(\xi)}{1 - \cos(\theta) \tanh(\xi)}, \\
k' &= k\big(\cosh(\xi) - \cos(\theta) \sinh(\xi)\big). 
}
The actions of parity and time reversal are
\eq{
	I_s f_{j m \lambda}(k) &= f_{j m -\lambda}(k)(-1)^j \\
	I_t f_{j m \lambda}(k) &= -f^*_{j -m \lambda}(k)(-1)^{j+m}.
}
\subsection{Angular momentum basis for irregular fields $|k j m \lambda \rangle^\text{\normalfont{in/out}}$}
Besides the regular angular momentum basis vectors, which are used to expand the incident field, the T-matrix formalism also uses the irregular outgoing angular momentum basis vectors to expand the scattered field. In the S-matrix formalism only irregular fields are used, to expand the incoming and outgoing fields. Physically, the energy flux of outgoing fields is outwards from the origin, while the flux is inwards towards the origin for the incoming fields. The regular fields have zero net flux. Mathematically, the difference between irregular and regular fields consists in the function responsible for the radial dependence: spherical Hankel functions of certain type for irregular fields instead of the spherical Bessel functions in Eq.~(\ref{eq:defR}) for regular fields. The Hankel functions have a singularity at $|\bm r|=0$. We define the incoming/outgoing states as:
\begin{equation}
   \boxed{
   \begin{aligned}
\ket{kjm\lambda}^\text{in/out}&\equiv\bm S^{\inout}_{j m\lambda}(k, \bm r, t) =\\
& \frac{1}{2}\sqrt{\frac{c\hbar}{\epsilon_0}} \frac{k \, e^{-i kc t}}{\sqrt{\pi}\sqrt{2j+1}}  \sum_{L=j-1}^{j+1} \sqrt{2L + 1}\, i^{L} h^{\inout}_{L}(k r) \, C^{j\lambda}_{L0,1\lambda} \bm Y^L_{j m}(\hat{\bm r}) \label{eq:defS}
   \end{aligned}
   }
\end{equation}
with spherical Hankel functions $h^{\inout}_{L} = j_{L} \mp i n_{L}$, $n_{L}$ being the spherical Neumann functions. 

The usual incoming/outgoing electric and magnetic multipoles are defined by substituting spherical Hankel functions of the second/first kind instead of the spherical Bessel functions in the regular multipoles in Eqs.(\ref{eq:regmultN} -\ref{eq:regmultM}):
\eq{
	\bm N^{\inout}_{jm}(k r, \hat{\bm r}) &= i  h^{\inout}_{j-1} (k r) \sqrt{\frac{j+1}{2j+1}} \bm Y^{j-1}_{j m}(\hat{\bm r})  -i  h^{\inout}_{j+1} (k r) \sqrt{\frac{j}{2j+1}} \bm Y^{j+1}_{j m}(\hat{\bm r}) \\
	\bm M^{\inout}_{jm}(k r, \hat{\bm r})  &= h^{\inout}_j (k r) \bm Y^j_{j m}(\hat{\bm r}).
}

The $\ket{kjm\lambda}^{\text{in/out}}$ can then also be written as:
\eq{
	\boxed{
		\bm S^{\inout}_{j m\lambda}(k, \bm r, t) = - \frac{1}{2} \sqrt{\frac{c\hbar}{\epsilon_0}} \frac{1}{\sqrt{2\pi}} \, k \, i^j  \Big(  e^{-i kc t}\, \bm N^{\inout}_{jm}(k r, \hat{\bm r}) + \lambda \,e^{-i kc t} \, \bm M^{\inout}_{jm}(k r, \hat{\bm r} ) \Big)}\label{eq:relation2}.
}

We highlight that an extra factor of $1/2$ in this definition leads to 
\begin{equation}
	\bm S^{\inn}_{j m\lambda} + \bm S^{\out}_{j m\lambda} = \bm R_{j m\lambda},\label{eq:new}
\end{equation}
the motivation and significance of which will be explained in Sec.~\ref{sec:pulses}.

The fact that irregular basis states $\bm S^{\inout}_{j m\lambda}$ transform as the regular fields $\bm R_{j m\lambda}$ under spatial translations and rotations is known \cite{peterson1973}. It is evident from the definition in Eq.~(\ref{eq:defS}) that irregular basis states also transform as the regular basis states under time translation. Their behavior under parity and time reversal are discussed in App.~\ref{app:AMparity}. In the next subsection we show that the irregular $\ket{kjm\lambda}^{\text{in/out}}$ transform under Lorentz boosts as the regular $\ket{kjm\lambda}$, completing the picture of their transformations under all isometries of the Minkowski space-time. This result is necessary for properly connecting the T-matrix and S-matrix formalisms to the \Poincare group.

\subsubsection{Irregular fields transform under boosts as regular fields\label{sec:irrboost}} 

In this section we discuss the transformation law for irregular basis vectors $\bm S_{jm\lambda}(k, \bm r, t)$ corresponding to either incoming or outgoing basis states in Eq.~(\ref{eq:defS}). We show that they transform in the same way as the regular basis vectors  $\bm R_{jm\lambda}(k, \bm r, t)$, according to Eq.~(\ref{eq:useful}):

\eq{
	\bm S^{\inout}_{jm\lambda}&(k, \bm r, t) \mapsto  \frac{1}{2}\,\sqrt{2j+1} \sum_{j'=1}^{\infty}\sqrt{2j'+1} \int^{1}_{-1} 
 d(\cos\theta) \, d^j_{m\lambda}(\theta) \,d^{j'}_{m\lambda}(\theta') \, \bm S^{\inout}_{j'm\lambda}(k', \bm r, t)\label{eq:LBirregular} 
}

Since the general formula for boosting electromagnetic field is independent of its type, Eq.~(\ref{eq:boost}) implies that
\eq{
	\bm S^{\inout}_{jm\lambda}&(k, \bm r, t) \rightarrow \Big (\cosh(\xi) \mathds{1} + i\lambda \sinh(\xi) \bm e_z \times - \frac{\sinh^2(\xi) \bm e_z}{\cosh(\xi)+1} \bm e_z \cdot \Big) \bm S^{\inout}_{jm\lambda}(k, \tilde{\bm r}, \tilde t) \label{eq:boostxi}
}
and since Lorentz boosts in the z-direction constitute a one-parameter Lie group, it is enough to prove the following equality of derivatives w.r.t. $\xi$ at zero:
\eq{
&\partial_\xi \frac{1}{2}\sqrt{2j+1}\sum_{j'=1}^{\infty}\sqrt{2{j'}+1} \int^{1}_{-1}  
 d(\cos\theta)  \,d^j_{m\lambda}(\theta) \, d^{j'}_{m\lambda}(\theta') \, \bm S^{\inout}_{{j'}m\lambda}(k', \bm r, t) \Big|_{\xi=0}\nonumber\\
 &=\partial_\xi  \Big (\cosh(\xi) \mathds{1} + i\lambda \sinh(\xi) \bm e_z \times - \frac{\sinh^2(\xi) \bm e_z}{\cosh(\xi)+1} \bm e_z \cdot \Big) \bm S^{\inout}_{jm\lambda}(k, \tilde{\bm r}, \tilde t) \Big|_{\xi=0}. \label{eq:chizero}
}
The proof is also simplified by the fact that regular fields already satisfy this condition and the only difference consists in the functions responsible for the radial dependence (spherical Hankel functions instead of spherical Bessel functions). Lengthy but straightforward calculations allow one to re-write Eq.~(\ref{eq:chizero}) (and the analogous expression for regular fields as well)
by separating the radial and angular dependencies of both sides as
\eq{
	 &r h^{\inout}_0(r) \bm A(\hat{\bm r}, t) + r h^{\inout}_1(r) \bm B(\hat{\bm r}, t) + \sum_{l=0}^N  h^{\inout}_l(r) \bm C_l(\hat{\bm r}, t)\nonumber \\
	 &= r h^{\inout}_0(r) \bm A'(\hat{\bm r}, t) + r h^{\inout}_1(r) \bm B'(\hat{\bm r}, t) + \sum_{l=0}^N h^{\inout}_l(r) \bm C'_l(\hat{\bm r}, t),\label{eq:hankels}
}
where the decomposition with primed coefficient functions corresponds to the right hand side of Eq.~(\ref{eq:chizero}) and the unprimed one to the left hand side. $N$ is finite, and for readability and without loss of generality we set $k = 1$. 

We use the fact that the statement in question already holds for regular fields, which means that exactly the same coefficients solve the equation for spherical Bessel functions:
\eq{
	 &r j_0(r) \bm A(\hat{\bm r}, t) + r j_1(r) \bm B(\hat{\bm r}, t) + \sum_{l=0}^N  j_l(r) \bm C_l(\hat{\bm r}, t)\nonumber \\
	 &= r j_0(r) \bm A'(\hat{\bm r}, t) + r j_1(r) \bm B'(\hat{\bm r}, t) + \sum_{l=0}^N j_l(r) \bm C'_l(\hat{\bm r}, t).
}
Writing
\eq{
	& \sin(r) \bm A(\hat{\bm r}, t) +\Big(\frac{\sin(r)}{r} - \cos(r) \Big)\bm B(\hat{\bm r}, t) + \sum_{l=0}^N j_l(r) \bm C_l(\hat{\bm r}, t) \nonumber\\
	 &= \sin(r) \bm A'(\hat{\bm r}, t) + \Big(\frac{\sin(r)}{r} - \cos(r) \Big) \bm B'(\hat{\bm r}, t) + \sum_{l=0}^N j_l(r) \bm C'_l(\hat{\bm r}, t)
}
one notes that in the limit $r \rightarrow \infty$ the spherical Bessel functions and $\frac{\sin( r)}{r}$ vanish, hence the coefficients at $\sin$ and $\cos$ functions must be equal: $\bm A(\hat{\bm r}, t) = \bm A'(\hat{\bm r}, t)$ and $\bm B(\hat{\bm r}, t) = \bm B'(\hat{\bm r}, t)$. Then, from
\eq{
	 \sum_{l=0}^N j_l(r) \bm C_l(\hat{\bm r}, t) &=  \sum_{l=0}^N j_l(r) \bm C'_l(\hat{\bm r}, t)
}
and the orthogonality of spherical Bessel functions follows the equality $\bm C_l(\hat{\bm r}, t) = \bm C'_l(\hat{\bm r}, t)$ for all $k$.
This proves the statement for the Hankel functions Eq.~(\ref{eq:hankels}), as well as for any functions that satisfy the same differential equation as spherical Bessel functions.

\subsubsection{Relation between incoming, outgoing, and regular fields} \label{sec:pulses}
Consider a regular electromagnetic pulse with a Gaussian profile of width $\Delta$ in $k=\omega/c$
\eq{
	\bm E_{\text{p}}(\bm r, t) &=  A \int_0^{\infty} dk \, k \,  e^{-\frac{(k-k_0)^2}{2\Delta^2} } \bm R_{jm\lambda}(k, \bm r, t), \label{eq:p1}
}
normalized with some constant $A$, and that is constructed as a spectral superposition of regular basis vector fields $\bm R_{jm\lambda}(k, \bm r, t)$ with some fixed $j$, $m$ and $\lambda$. It can be decomposed into incoming and outgoing parts using the connection between spherical Bessel and Hankel functions $j = (h^{\inn} + h^{\out})/2$ as
\eq{
	\bm E_{\text{p}}(\bm r, t) &=  \bm E^{\text{in}}_{\text{p}}(\bm r, t) + \bm E^{\text{out}}_{\text{p}}(\bm r, t) \label{eq:regtoincandout}
} 
with
\eq{
 \bm E^{\text{in}}_{\text{p}}(\bm r, t) &= A \int_0^{\infty} dk \, k \,  e^{-\frac{(k-k_0)^2}{2\Delta^2} } \bm S^{\inn}_{jm\lambda}(k, \bm r, t) \label{eq:ingg}\\
 \bm E^{\text{out}}_{\text{p}}(\bm r, t)&= A \int_0^{\infty} dk \, k \,  e^{-\frac{(k-k_0)^2}{2\Delta^2} } \bm S^{\out}_{jm\lambda}(k, \bm r, t)\label{eq:outg}
}
for $\abs{\bm r} > 0$.

The regular pulse $\bm E_{p}(\bm r, t)$ has finite length, so it is possible to define a time period before the pulse first reaches the origin at $\bm r=\bm 0$, and a time period after the pulse has completely crossed the origin. During the first period the pulse completely consists of the incoming part 
\eq{
\bm E_{\text{p}}(\bm r, t) &= \bm E^{\text{in}}_{\text{p}}(\bm r, t) 
} 
while $\bm E^{\text{out}}_{\text{p}}(\bm r, t) = 0$, and during the second period it solely consists of the outgoing part
\eq{
\bm E_{\text{p}}(\bm r, t) &= \bm E^{\text{out}}_{\text{p}}(\bm r, t) 
}
while $\bm E^{\text{in}}_{\text{p}}(\bm r, t) = 0$. 

\begin{figure}[h]
	\centering
        \includegraphics[width=\textwidth]{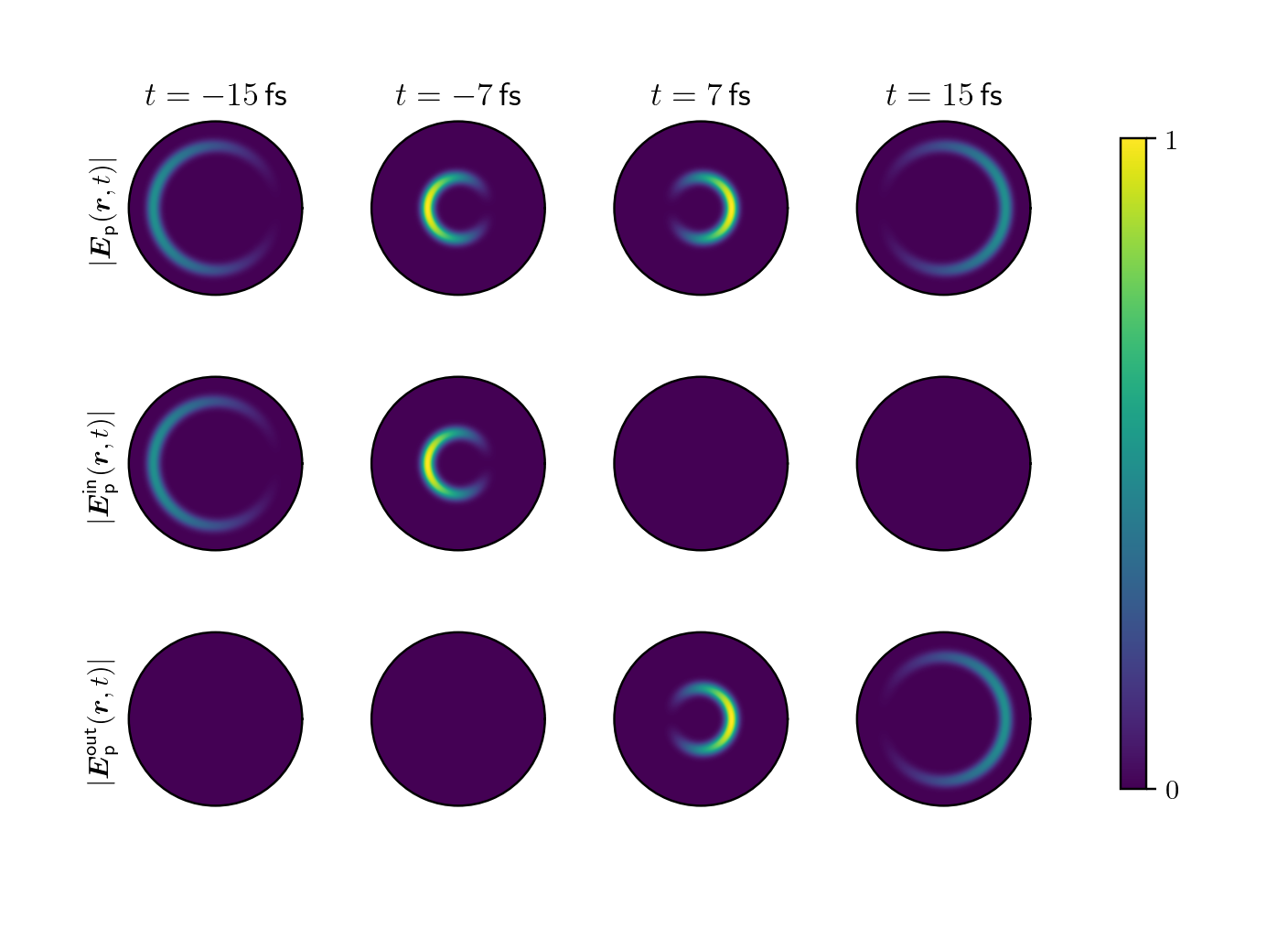}
	\caption{Comparison of Gaussian pulses with center wavelength $\frac{2\pi}{k_0} = 400~\si{nm}$ and Gaussian width $\Delta^{-1} = 300~\si{nm}$ constructed as spectral superposition {\em with the same coefficients} of basis fields of different types: regular $\bm R_{jm\lambda}(k, \bm r, t)$ (top), incoming $\bm S^{\inn}_{jm\lambda}(k, \bm r, t)$ (middle) and outgoing $\bm S^{\out}_{jm\lambda}(k, \bm r, t)$ (bottom), for $j = 1$, $m=1$ and helicity $\lambda = 1$. Figures depict absolute values of the electric field in the $xz$-plane at four different points in time. The incoming pulse is computed to be identically zero after the end of its absorption in the origin, and is equal to the corresponding regular field for times before the start of its absorption. The outgoing pulse is computed to be  identically zero before the start of its emission from the origin, and is equal to the corresponding regular field for times after the end of its emission.
	\label{fig:Pulse}}
\end{figure}

A representative example is shown on Fig.~\ref{fig:Pulse}. We plot numerically computed values of concrete Gaussian pulses Eqs.~(\ref{eq:p1}),(\ref{eq:ingg}),(\ref{eq:outg}) with total angular momentum $j=1$, angular momentum around $z$-axis $m=1$ and helicity $\lambda = 1$. The time stamps are selected to be within the two periods defined above, when irregular pulses are either equal to the regular pulse or identically zero. In these periods, the substitution of irregular basis fields by regular basis fields would not change the value of the total field of the pulse, which can be useful for practical applications because the spherical Bessel functions in regular fields are numerically better behaved than the spherical Hankel functions \cite[App.~B]{Garcia2018}. The notable split illustrated in Fig.~\ref{fig:Pulse} does not happen for monochromatic fields, e.g. when beams of infinite duration are involved, because at each point of time and space the regular field contains contributions from both incoming and outgoing components. 

In general, such connection is present in regions of space-time when the regular field is known to contain only an incoming or only an outgoing part. 

Now, if we consider coefficients $f_{jm\lambda}(k)$ that have finite norm $\braket{f|f}$, then, when combined with regular basis fields $\bm R$, they describe a regular freely propagating physical field. When the same coefficients are combined with outgoing basis fields $\bm S^{\out}$, they describe an emitted electromagnetic field that is zero before the start of the emission and that is identically equal to the corresponding regular field at times after the end of the emission. On the other hand, when  $f_{jm\lambda}(k)$  are combined with incoming basis fields $\bm S^{\inn}$, the linear combination will result in an electromagnetic field that will be absorbed during some time, such that after this period the field will be zero and before the start of the absorption the field will be identically equal to the regular field combined with the same coefficients. This allows one to connect irregular fields to the Hilbert space formalism and to use the same scalar product Eqs.~(\ref{eq:scalarPW}),(\ref{eq:scalarAM}) to, in particular, compute quantities of emitted or absorbed fields, such as energy and momentum. While using the ket notation, we will distinguish incoming and outgoing types of fields that share the same coefficients $f_{jm\lambda}(k)$ by a superscript $\ket{f}^\text{in}$ or $\ket{f}^\text{out}$, versus the regular $\ket{f}$. 

We can now see that our definition of basis fields $\bm S^{\inout}$ that incorporates an extra factor of $1/2$ when compared to its regular counterpart, and that differs from the usual approaches, follows from requiring the use of the same scalar product for regular, incoming and outgoing fields. Let us consider the number of photons in a regular pulse, which does not change with time. The decomposition of the regular pulse as $\ket{f} = \ket{f}^\text{in} + \ket{f}^\text{out}$ applies at all times, only that there are time periods where either $\ket{f}^\text{out}$ or  $\ket{f}^\text{in}$ vanish. Notably, the total number of photons in the regular pulse $\braket{f|f}$ is equal to the number of photons absorbed~$\fourIdx{\text{in}}{}{\text{in}}{}{\braket{f|f}}$ and also equal to the number of photons emitted~$\fourIdx{\text{out}}{}{\text{out}}{}{\braket{f|f}}$.

\subsection{On the convergence regions of some expansions}
A discussion about the validity of particular kinds of field expansions is in order at this point. In the previous sections, we have identified the coefficient functions in expansions of electromagnetic fields as members of the Hilbert space. We argue here that, while some of those expansions do not converge at all space-time points, where more complicated expansions are needed, the coefficient functions contain sufficient information to recover the fields at all points outside material objects. The T-matrix and the S-matrix that we discuss in the next section are linear mappings between such coefficient functions.

Irregular basis fields are not defined in the whole space because they are singular at the origin $\abs{\bm r} = 0$. However, in \Eq{eq:regtoincandout}, the validity of the separation of a general regular pulse into the incoming and outgoing parts $\bm E^{\text{in}}_{\text{p}}(\bm r, t)$ and $\bm E^{\text{out}}_{\text{p}}(\bm r, t)$ depends on the pulse and may be restricted by a stronger condition than $\abs{\bm r} > 0$. For example, a spatial translation of the pulse \Eq{eq:regtoincandout} by some distance $a$ in any direction would decrease the region of validity of Eqs.~(\ref{eq:ingg}-\ref{eq:outg}) to points outside of the sphere with radius $a$, $\abs{\bm r} > a$, while inside such sphere one must use another expansion branch featuring regular fields \cite[Eq.(47ab)]{wittmann1988}. Similarly, the Lorentz boosted irregular electromagnetic field in \Eq{eq:LBirregular} will not converge for all the $(\bm r,t)$ points, and an expansion with branches would also be then needed. A similar issue that arises in the T-matrix formalism is the validity of the expansion of the scattered field into only outgoing multipoles, similar to \Eq{eq:outg}, which is strictly valid only outside the smallest sphere enclosing the scatterer. This issue, which can be addressed with more complicated expansions \cite{Theobald2017,Egel2017,Martin2019,Schebarchov2019,Lamprianidis2023}, also affects the S-matrix formalism, and the latter could also need branches to expand the incoming field, as in the case of the spatially translated pulse that we just discussed. However, even though expansions with branches are sometimes needed, it should be noted that the expansion coefficients of the far field are sufficient information to determine the field everywhere outside material objects. This follows from the fact that such coefficients determine the far field, which at its turn determines the field everywhere outside material objects \cite[Theorems~6.9~and~6.10]{Colton2012}. The work in \cite{Martin2019} provides a clear illustration of this, since an accurate T-matrix of two nearby disks that grossly invade each other's smallest enclosing spheres can be computed using solely the positions and the T-matrices of each individual disk. Therefore, the expansion coefficient functions that constitute the Hilbert space, which determine the far fields, are sufficient information to recover the fields everywhere outside material objects, albeit potentially through relatively complicated expansions with e.g. several branches.

\section{Polychromatic T-matrix and  S-matrix }\label{sec:ST}
We now have all the necessary elements in place for defining the polychromatic T-matrix. Afterwards we will define the polychromatic S-matrix. Both operators contain the same information and are bijectively connected, but they map different parts of the total electromagnetic field. The T-matrix connects the regular field, called incident field, with the irregular outgoing field, called scattered field. The S-matrix connects irregular incoming fields to irregular outgoing fields. 

\begin{figure}[h]
	\begin{center}
	\begin{overpic}[width=0.75\linewidth]{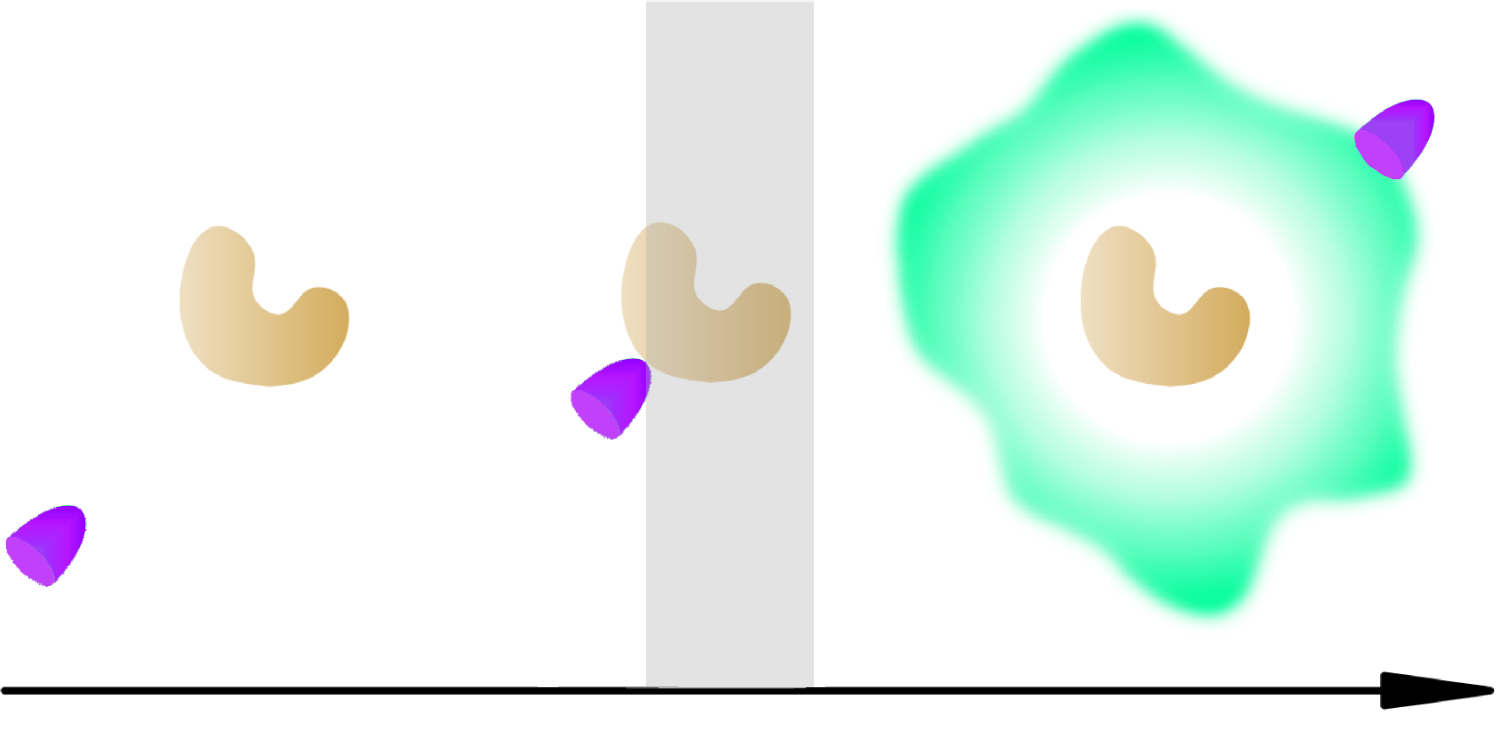}
		\put(42, -0.3){$t_1$}
		\put(52, -0.3){$t_2$}
		\put(92,5){time}
	\end{overpic}
	\end{center}
		\caption{An electromagnetic pulse interacts with a material object during a finite time (gray shade). The $T$-matrix is a linear operator in the Hilbert space of solutions of Maxwell equations that maps the incident fields (lilac) to the scattered fields (green). The light-matter interaction starts at $t=t_1$. Before $t_1$, causality forbids the existence of any scattered field. Time $t=t_2$ marks the end of the emission of the scattered field from the object. \label{fig:lmi}}
\end{figure}

Let us start by considering the light-matter interaction picture in Fig.~\ref{fig:lmi}, where a light pulse interacts with a material object.

\subsection{Polychromatic T-matrix} \label{sec:Tpol}
In the usual definition of the monochromatic T-matrix \cite[Sec. 5.1]{mishchenko2002}, a time-harmonic field outside of the smallest sphere that encloses a localized electromagnetic scatterer can be written as a sum of the regular and outgoing multipoles
\eq{
	\bm E(k, \bm r, t) &= e^{-ikct} \sum_{j=1}^{\infty}\sum_{m=-j}^{j}\, a_{j m} \, \bm N_{jm}(k r, \hat{ \bm{r}})   +  b_{j m} \, \bm M_{jm}(k r, \hat{ \bm{r}})   \nonumber \\
	&+ e^{-ikc t} \sum_{j=1}^{\infty}\sum_{m=-j}^{j} \, p_{j m} \, \bm N^{\out}_{jm}(k r, \hat{ \bm{r}})   +  q_{j m} \, \bm M^{\out}_{jm}(k r, \hat{ \bm{r}}) \label{eq:FieldIncScatMono}
}
with the second equation valid for $r = |\bm r|$ larger than the radius of the smallest sphere enclosing the object. The first, regular part of \Eq{eq:FieldIncScatMono} is called the incident field and the second, irregular part is called the scattered field. 

In the case of a single monochromatic field, the usual monochromatic T-matrix is defined as the matrix that maps the coefficients of the incident and the scattered electromagnetic fields:
\eq{
	\begin{pmatrix}
		\vec p \\ \vec q
	\end{pmatrix}
	=
	T_\text{u}
	\begin{pmatrix}
		\vec a \\ \vec b
	\end{pmatrix}.	\label{eq:TmatMono}
}

However, the most general linear scattering situation concerns interaction of a polychromatic field with an object. A total field in this case is a spectral superposition of monochromatic fields: 
\eq{
	\bm E(k, \bm r, t) &= \int_0^{\infty} dk \, e^{-ikct} \sum_{j=1}^{\infty}\sum_{m=-j}^{j}\, a_{j m}(k) \, \bm N_{jm}(k r, \hat{ \bm{r}})   +  b_{j m}(k) \, \bm M_{jm}(k r, \hat{ \bm{r}})   \nonumber \\
	&+ \int_0^{\infty} dk \,  e^{-ikc t} \sum_{j=1}^{\infty}\sum_{m=-j}^{j} \, p_{j m}(k) \, \bm N^{\out}_{jm}(k r, \hat{ \bm{r}})   +  q_{j m}(k) \, \bm M^{\out}_{jm}(k r, \hat{ \bm{r}}). \label{eq:FieldIncScatPoly}
}
A principal difference to the monochromatic picture consists in the fact that a general linear connection between the incident and scattered field allows coupling of different frequencies. An example of a physical situation when this coupling is necessary is the relativistic scattering: a monochromatic beam that hits a moving object will produce a scattered field with components of several different frequencies.

Following the path suggested by the representation theory, we connect the electric field to the Hilbert space of solutions of Maxwell's equations by writing it in terms of the basis fields $\bm R_{jm\lambda}(k)$ and $\bm S^{\out}_{jm\lambda}(k)$:
\eq{
	\bm E(\bm r, t) &=  \int_0^\infty dk \, k  \sum_{\lambda=\pm1}\sum_{j=1}^{\infty}\sum_{m=-j}^{j} \, f_{j m\lambda}(k) \, \bm R_{jm\lambda}(k,\bm r, t) \nonumber \\
	&+  \int_0^\infty dk \, k  \sum_{\lambda=\pm1}\sum_{j=1}^{\infty}\sum_{m=-j}^{j} \, g_{j m\lambda}(k)\, \bm S^{\out}_{jm\lambda}(k, \bm r, t),
\label{eq:totpolyT}
}
or, equivalently  

\begin{equation}
   \boxed{
   \begin{aligned}
	   \ket{f} + \ket{g}^\text{out}&= \int_0^\infty dk \, k\, \sum_{\lambda=\pm1} \sum_{j=1}^{\infty}\sum_{m=-j}^{j}\, f_{j m\lambda}(k) \, \ket{k j m \lambda}  \label{eq:f+g}\\
	   &+\int_0^\infty dk \, k\, \sum_{\lambda=\pm1} \sum_{j=1}^{\infty}\sum_{m=-j}^{j}\, g_{j m\lambda}(k)  \,\ket{k j m \lambda}^\text{out},
   \end{aligned}
}
\end{equation}
	with coefficients that follow from Eqs.(\ref{eq:totpolyT}), (\ref{eq:relation}), (\ref{eq:relation2}), and that, crucially, are compatible with the scalar product in \Eq{eq:scalarAM}:
\begin{equation}
   \boxed{
   \begin{aligned}
	   &f_{j m \lambda}(k) = -  \sqrt{\frac{\epsilon_0 }{c\hbar}}\frac{\sqrt{2\pi}(-i)^j}{k^2} \big( a_{jm}(k) + \lambda b_{jm}(k) \big) \\
	   &g_{j m \lambda}(k) = -  \sqrt{\frac{\epsilon_0 }{c\hbar}}\frac{\sqrt{2\pi}(-i)^j}{k^2} \big( p_{jm}(k) + \lambda q_{jm}(k) \big).\\
   \end{aligned}
   }\label{eq:fabgpq}
\end{equation}

In linear light-matter interactions the coefficients of the scattered field $g_{j m \lambda}(k)$ are linearly related to the coefficients of the incident field $f_{j m \lambda}(k)$. We define the polychromatic T-matrix as the linear operator mapping the regular incident field to the outgoing scattered field via
\eq{
	 \ket{g}^\text{out} &= T \ket{f}
	 \label{eq:Tinout}
}
which implies for the coefficients
\begin{equation}
   \boxed{
   \begin{aligned}
	   \label{eq:Tnew}
	g_{j_1m_1\lambda_1}(k_1)& = \int_0^\infty dk_2 \, k_2 \sum_{\lambda_2=\pm 1} \sum_{j_2=1}^{\infty}\sum_{m_2=-j}^{j} \, T^{j_1m_1\lambda_1}_{\,j_2m_2\lambda_2}(k_1, k_2) f_{j_2m_2\lambda_2}(k_2),\\
	\text{ where }&T^{j_1m_1\lambda_1}_{\,j_2m_2\lambda_2}(k_1, k_2) = \fourIdx{\text{out}}{}{}{}{\!\braket{ k_1 j_1 m_1 \lambda_1|T|k_2 j_2 m_2 \lambda_2} } .
\end{aligned}
   }
\end{equation}

\subsection{Building frequency-diagonal polychromatic T-matrices from monochromatic T-matrices \label{sec:diagonal}}
Quite often, one considers scattering processes where frequencies do not change during light-matter interaction. Such processes can hence be described by T-matrices that are diagonal in frequency. Here we show how T-matrices that do not mix frequency are a special case of the polychromatic T-matrix, and provide the formula for building the polychromatic T-matrix from the usual monochromatic T-matrices. 

Consider scattering of an incident field
\eq{
\bm E_{\text{inc}}(\bm r, t) &=  \int dk \, e^{-ikc t} \sum_{j=1}^{\infty}\sum_{m=-j}^{j} \, \bm N_{jm}(k r, \hat{ \bm{r}})  \, a_{j m}(k) + \bm M_{jm}(k r, \hat{ \bm{r}})  \, b_{j m}(k) \label{eq:inc-same}
}
to a scattered field
\eq{
\bm E_{\text{sc}}(\bm r, t) &=  \int dk \,  e^{-ikc t} \sum_{j=1}^{\infty}\sum_{m=-j}^{j} \, \bm N^{\out}_{jm}(k r, \hat{ \bm{r}})  \, p_{j m}(k) + \bm M^{\out}_{jm}(k r, \hat{ \bm{r}}) \, q_{j m}(k),\label{eq:sc-same}
}
where the usual monochromatic T-matrices connect the coefficients at each frequency $\omega = kc$ as 
\eq{
	\begin{pmatrix} \vec p(k) \\ \vec q(k) \end{pmatrix} = T_{u}(k) \begin{pmatrix}\vec a(k) \\ \vec b(k) \end{pmatrix} = \begin{pmatrix} T^{NN}_u(k) & T^{NM}_u(k) \\T^{ME}_u(k) &T^{MM}_u(k) \end{pmatrix} \begin{pmatrix} \vec a(k) \\ \vec b(k) \end{pmatrix}.
}
According to Eq.~(\ref{eq:Tnew}), the same scattering is realized by the following polychromatic T-matrix, which is diagonal in frequency (and written in the helicity basis):
{\small
\begin{equation}
   \boxed{
   \begin{aligned}
	&T^{j_1 m_1 \lambda_1}_{j_2 m_2 \lambda_2}(k_1, k_2) =\\
	   &\frac{1}{k_2}\delta(k_1 - k_2) \Big (T^{NN}_u(k_2)^{j_1 m_1}_{j_2 m_2 } + \lambda_1 T^{MN}_u(k_2)^{j_1 m_1}_{j_2 m_2 } + \lambda_2 T^{NM}_u(k_2)^{j_1 m_1}_{j_2 m_2 } + \lambda_1 \lambda_2 T^{MM}_u(k_2)^{j_1 m_1}_{j_2 m_2 } \Big)\label{eq:TTu}
   \end{aligned}
}
\end{equation}
}
Equation~(\ref{eq:TTu}) follows from the decomposition of the fields in Eqs.~(\ref{eq:inc-same} - \ref{eq:sc-same}), and from \Eq{eq:fabgpq}.

With the contents of this section, the monochromatic T-matrices $T_u(k)$ computed with the usual conventions by currently available formulas and computer codes can be easily re-used for computing the polychromatic T-matrix in the new conventions.

\subsection{Polychromatic S-matrix\label{sec:Spol}}
An equivalent description of scattering may be provided by the S-matrix formalism, which is based on the decomposition of the total electromagnetic field into the incoming and the outgoing fields \cite[Eq.~(5.47)]{mishchenko2002}. Again, we generalize the monochromatic setting by considering the total field as the spectral superposition of monochromatic fields 
\eq{
	\bm E(k, \bm r, t) &= \int_0^\infty dk \,  e^{-ikct} \sum_{j=1}^{\infty}\sum_{m=-j}^{j}\, a_{j m}(k) \, \bm N^{\inn}_{jm}(k r, \hat{ \bm{r}})   +  b_{j m}(k) \, \bm M^{\inn}_{jm}(k r, \hat{ \bm{r}})   \nonumber \\
	&+ \int_0^\infty dk \, e^{-ikc t} \sum_{j=1}^{\infty}\sum_{m=-j}^{j} \, \rho_{j m}(k) \, \bm N^{\out}_{jm}(k r, \hat{ \bm{r}})   +  \mu_{j m}(k) \, \bm M^{\out}_{jm}(k r, \hat{ \bm{r}}). \label{eq:InOutFields}
}

Similarly to the previous section, we proceed by writing the total field Eq.~(\ref{eq:InOutFields}) in terms of $\bm S^{\inout}_{j m\lambda}$
\eq{
	\bm E(k, \bm r, t) &= \int_0^\infty dk \, k\, \sum_{\lambda=\pm1} \sum_{j=1}^{\infty}\sum_{m=-j}^{j}\, f_{j m\lambda}(k) \, \bm S^{\inn}_{j m\lambda}(k, \bm r, t)   \nonumber \\
	&+ \int_0^\infty dk \, k\, \sum_{\lambda=\pm1} \sum_{j=1}^{\infty}\sum_{m=-j}^{j}\, h_{j m\lambda}(k)  \,\bm S^{\out}_{j m\lambda}(k, \bm r, t) , \label{eq:InOutFieldsGT} 
	}
or, equivalently  

\begin{equation}
   \boxed{
   \begin{aligned}
	   &\ket{f}^\text{in} + \ket{h}^\text{out}= \int_0^\infty dk \, k\, \sum_{\lambda=\pm1} \sum_{j=1}^{\infty}\sum_{m=-j}^{j}\, f_{j m\lambda}(k) \, \ket{k j m \lambda}^\text{in}  \label{eq:fh}\\
	   &\hspace{3cm}+\int_0^\infty dk \, k\, \sum_{\lambda=\pm1} \sum_{j=1}^{\infty}\sum_{m=-j}^{j}\, h_{j m\lambda}(k)  \,\ket{k j m \lambda}^\text{out},
   \end{aligned}
}
\end{equation}
	with coefficients that follow from \Eq{eq:InOutFields} and \Eq{eq:relation2}, and that, crucially, are compatible with the scalar product in \Eq{eq:scalarAM}:
\begin{equation}
   \boxed{
   \begin{aligned}
	   &f_{j m \lambda}(k) = -  \sqrt{\frac{\epsilon_0 }{c\hbar}}\frac{\sqrt{2\pi}(-i)^j}{k^2} \big( a_{jm}(k) + \lambda b_{jm}(k) \big) \\
	   &h_{j m \lambda}(k) = -  \sqrt{\frac{\epsilon_0 }{c\hbar}}\frac{\sqrt{2\pi}(-i)^j}{k^2} \big( \rho_{jm}(k) + \lambda \mu_{jm}(k) \big).\\
   \end{aligned}
   }
\end{equation}

In linear light-matter interactions, the coefficients of the outgoing field $h_{j m \lambda}(k)$ are linearly related to the coefficients of the incoming field $f_{j m \lambda}(k)$. We define the polychromatic S-matrix as the linear operator mapping the incoming field to the outgoing field via
\eq{
	 \ket{h}^\text{out} &= S \ket{f}^\text{in}
	 \label{eq:Sinout}
}
which implies for the coefficients
\begin{equation}
   \boxed{
   \begin{aligned}
	   \label{eq:Soperation}
	h_{j_1m_1\lambda_1}(k_1)& = \int_0^\infty dk_2 \, k_2 \sum_{\lambda_2=\pm 1} \sum_{j_2=1}^{\infty}\sum_{m_2=-j}^{j} \, S^{j_1m_1\lambda_1}_{\,j_2m_2\lambda_2}(k_1, k_2) f_{j_2m_2\lambda_2}(k_2),\\
	\text{ where }&S^{j_1m_1\lambda_1}_{\,j_2m_2\lambda_2}(k_1, k_2) =  \fourIdx{\text{out}}{}{\text{in}}{}{\!\braket{k_1 j_1 m_1 \lambda_1|S|k_2 j_2 m_2 \lambda_2} } .
\end{aligned}
}
\end{equation}

The connection between the T-matrix and the S-matrix formalisms is determined by decompositions \Eq{eq:totpolyT} and \Eq{eq:InOutFieldsGT}. We start with the T-matrix decomposition of total field into the incident and the scattered components and separate the regular part into the incoming and scattered fileds
\eq{
\ket{f} + \ket{g}^\text{out} &= \ket{f}^\text{in} + \ket{f}^\text{out} + \ket{g}^\text{out}. 
}
This brings the total field to the decomposition that underlines the S-matrix formalism, implying
\eq{
	S  \ket{f}^\text{in} &= \ket{f}^\text{out} + \ket{g}^\text{out} := \ket{h}^\text{out}, \label{eq:ffgh}
}
while the T-matrix maps the fields as
\eq{
	T \ket{f} &= \ket{g}^\text{out}.\label{eq:tfg}
}
We see that in our convention the coefficients of the incident and of the incoming field are equal, namely $f_{jm\lambda}(k)$, and the coefficients of the scattered field are connected to the coefficients of the incident and of the outgoing fields via $h_{jm\lambda}(k) = f_{jm\lambda}(k) + g_{jm\lambda}(k)$.

Equation~(\ref{eq:ffgh}) and \Eq{eq:tfg} imply the connection between the T-matrix and the S-matrix as operators to be
\begin{equation}
\boxed{
\begin{aligned}
	S \ket{f}^\text{in} = \ket{f}^\text{out} + T \ket{f} \label{eq:operatorconn}
\end{aligned}
}
\end{equation}
for arbitrary coefficients $f_{jm\lambda}(k)$. Numerically, when the elements of the T-matrix are known, the elements of the S-matrix may be computed via the simple relation
\eq{
	S = \mathds{1} + T, \label{eq:s1t}
}
since the distinction between the incident, incoming and outgoing fields plays a role only when combining the coefficients with the corresponding basis elements to construct the physical field $\bm E(\bm r, t)$. 

\Eq{eq:s1t} also allows to straightforwardly obtain results identical to of Sec.~\ref{sec:diagonal} for the case of the frequency-diagonal S-matrix.

We note that the connection in \Eq{eq:s1t} differs from a more common formula
\eq{
	S_\text{u} = \mathds{1} + 2T_\text{u}, \label{eq:s1told}
} 
where the subscript 'u' stands for the usual way of defining basis states. The reason lies in the way our basis fields $\bm S^{\inout}_{j m\lambda}$ are defined, namely via substituting spherical Bessel functions $j_L(kr)$ in $\bm R_{j m\lambda}$ by spherical Hankel functions halved $h^{\inout}_L(kr)/2$. In the usual approach, however, the irregular vector spherical functions $ \bm M^{\inout}_{jm}$ and $ \bm N^{\inout}_{jm}$ are defined by substituting spherical spherical Bessel functions by spherical Hankel functions without division by 2. 

While the T-matrix in our convention is larger than the usual one by the factor of two, the S-matrix is identical in both conventions, because incoming and outgoing basis fields have been changed in the same way, and therefore numerical values of the S-matrix elements  \fourIdx{\text{out}}{}{\text{in}}{}{\!\braket{k_1 j_1 m_1 \lambda_1|S|k_2 j_2 m_2 \lambda_2} } do not change.

\section{Transfer of energy and momentum from a light pulse to a Si sphere \label{sec:Transfer}}
The scalar product in Eq.~(\ref{eq:scalarPW}) or Eq.~(\ref{eq:scalarAM}) allows one to compute fundamental physical quantities, such as energy, momentum, and angular momentum carried by electromagnetic field $\ket{f}$:
\eq{
\braket{\Gamma} = \braket{f |  \Gamma | f}, \label{eq:quantity}
}
where $\Gamma$ is the Hermitian operator of the corresponding physical quantity: generator of time translations $c P^0=H$ for energy, generators for spatial translations $P_\alpha$ ($\alpha=x,y,z$) for linear momentum, and generators of rotations $J_\alpha$ ($\alpha=x,y,z$)  for angular momentum. If the scattering process is subject to a conservation law, then the difference between the quantities contained in incoming and outgoing fields is equal to the amount of the quantity transferred to or extracted from the object. For the purpose of computing this difference it is most convenient to describe scattering in terms of the S-matrix. Given the incoming field $\ket{f}$, the transferred amount $\braket{\Delta \Gamma}$ is \cite[Eq.~(3)]{fernandez2017}
\eq{
\braket{\Delta \Gamma} &= \fourIdx{\text{in}}{}{\text{in}}{}{\!\braket{f | \Gamma | f} } - \fourIdx{\text{out}}{}{\text{out}}{}{\!\braket{h | \Gamma | h} } \nonumber \\
&= \fourIdx{\text{in}}{}{\text{in}}{}{\!\braket{f | \Gamma - S^\dagger \Gamma S  | f} } , 
}
or in terms of the T-matrix 
\eq{
	\braket{\Delta \Gamma} &= -\braket{f | \, \Gamma T +  T^\dagger \Gamma +  T^\dagger \Gamma T \, | f} \label{eq:Ttransfer}
}
for a given incident field $\ket{f}$.

We illustrate the transfer of quantities with a left-handed ($\lambda=+1$) circularly polarized pulse with Gaussian profiles in time and space, described by the wave function at positive $\cos\theta$ as
\eq{
f_+(\bm k) &= A\, e^{i\phi } \cos\theta (1+\cos\theta) \,e^{- (k - k_0)^2 \Delta_t^2 c^2 / 2} \, e^{- k^2(1-\cos^2\theta) \Delta_\rho^2 / 2 },\\
f_-(\bm k) &= 0,
}
and set $f_\lambda(\bm k) = 0$ for $\cos\theta < 0$. The angles $\theta$ and $\phi$  are the polar and azimuthal angles of $\bm k$ respectively. We choose $\Delta_t = 10~\si{\fs}$, width $\Delta_\rho = 1~\si{\micro\m}$, the central wavelength  $\frac{2\pi}{k_0} = 380~\si{\nm}$, and set the constant $A = 3.8\times10^{8}~\si{\per\m}$ to fix the energy of the pulse to 1~\si{mJ}:
\eq{
	\braket {f|H|f} = \sum_{\lambda=\pm 1} \int \frac{d^3 \bm k}{k} \abs{f_\lambda(\bm k)}^2 c k = 1.0 \times 10^{-3}~\si{\J},
}
which is a concrete realization of \Eq{eq:quantity}. Similarly, the momentum in $z$-direction that is contained in the pulse can be computed via
\eq{
	\braket {f|P_z|f} =  \sum_{\lambda=\pm 1} \int \frac{d^3 \bm k}{k} \abs{f_\lambda(\bm k)}^2  \hbar k \cos\theta = 3.3 \times 10^{-12}~\si{\kg\m\per\s}.
}

Now let us consider interaction of the defined pulse with a silicon sphere of radius $100~\si{\nm}$ located in the origin of the reference frame. 

\begin{figure}[h]
 \begin{subfigure}[b]{0.5\textwidth}
        \centering
        \includegraphics[width=\textwidth]{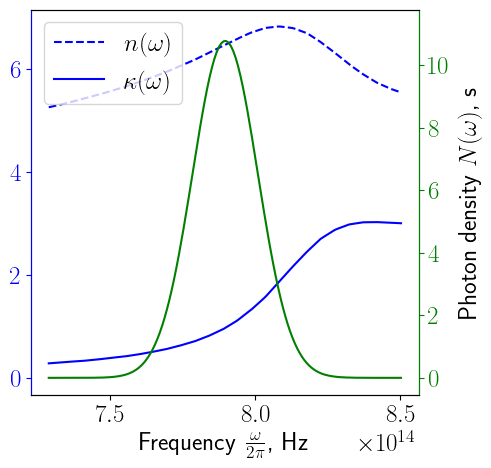}
        \caption{}
	\label{fig:SiliconPhotonDensity}
     \end{subfigure}
     \hfill
     \begin{subfigure}[b]{0.5\textwidth}
         \centering
        \includegraphics[width=\textwidth]{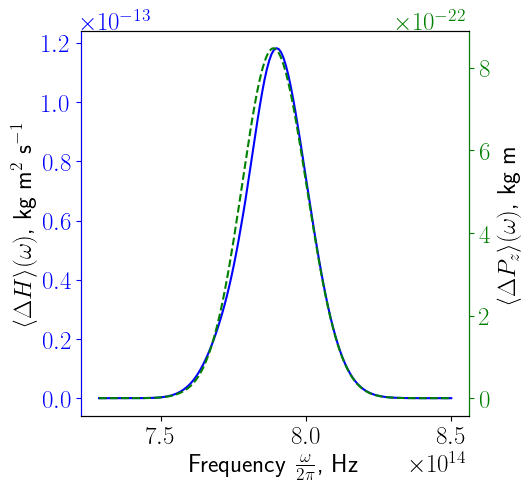}
        \caption{}
	\label{fig:Transfer}
     \end{subfigure}
	\caption{
		(a) Refractive index $n(\omega)$ and extinction coefficient $\kappa(\omega)$ of Silicon as a function of frequency (blue) together with the photon density with respect to the frequency of the incident field (green). (b) Transfer of energy from the pulse to the object per frequency (blue), together with the transfer of linear momentum in $z$-direction from the pulse to the object per frequency (green). Integrals of the functions provide the total transferred quantity.
		}
\end{figure}

The spectral content of the incident pulse in terms of its photon density w.r.t. angular frequency $\omega = k c$:
\eq{
N(\omega) = \frac{\omega}{c^2} \sum_{\lambda = \pm 1} \sum_{j=1}^{\infty} \sum_{m = -j}^j \abs{f_{jm\lambda}(\omega/c)}^2
}
is illustrated on top of the optical parameters of the silicon on the Fig.~\ref{fig:SiliconPhotonDensity}. 

To compute the transfer of the energy and of the momentum one requires the infinitesimal versions of the transformation laws Eq.~(\ref{eq:AMtimetr}-\ref{eq:AMztr}) in the angular-momentum basis, which read
\eq{
	&H f_{j m\lambda}(k) = \hbar c k f_{j m\lambda}(k) \\
	&P_z f_{j m\lambda}(k)= \nonumber\\
	&  \hbar k  \sqrt{2j+1} (-1)^{m-\lambda} \sum_{j' = j-1}^{j+1} \sqrt{2j'+1}    
	\begin{pmatrix}
	j & j' & 1 \\
	-m & m & 0
	\end{pmatrix} 
		\begin{pmatrix}
	j & j' & 1 \\
	-\lambda & \lambda & 0
	\end{pmatrix} 
	f_{j'm\lambda}(k),
}
where $\begin{pmatrix}
	j_1 & j_2 & j_3 \\
	m_1 & m_2 & m_3
	\end{pmatrix} $ are the Wigner 3-j symbols.
We generate the T-matrix of the sphere with the {\large \texttt{treams}} python package \cite{beutel2021,Beutel2023}, which is publicly available at \url{https://github.com/tfp-photonics/treams}, and use Eq.~(\ref{eq:Ttransfer}) to get the transfer of energy and momentum to the object: 
\eq{
	\braket{\Delta H} &= 2 \times 10^{-8}~\si{\J} \\
	\braket{\Delta P_z} &= 1.5 \times 10^{-16}~\si{\kg\m\per\s}.
}

A more specific information on the transfer is provided in Fig.~\ref{fig:Transfer}, where the density of transferred quantity with respect to the frequency is plotted for both energy $\braket{\Delta H} (\omega)$ and momentum $\braket{\Delta P_z} (\omega)$ transfer. The amounts of total transfer are connected to the corresponding densities as
\eq{
	\braket{\Delta H} &= \int_0^{\infty} d\omega \, \braket{\Delta H} (\omega)\\
	\braket{\Delta P_z} &= \int_0^{\infty} d\omega \, \braket{\Delta P_z} (\omega) .
}

This method differs from alternative approaches that make use of the Maxwell's stress tensor \cite{preez2015}. Instead, we employ the scalar product formula \Eq{eq:scalarAM}, which enables us to achieve accurate results while highlighting the underlying group-theoretical principles.

\section{Conclusions}

In this work, we have generalised the T-matrix method to the polychromatic setting by exploiting the connection between electromagnetism and the theory of group representations. This extension broadens the range of scenarios that can be accurately modelled and studied using the T-matrix method, allowing for a more comprehensive understanding of electromagnetic scattering phenomena.

Through the introduction of a novel convention for electromagnetic basis fields, which possess the distinctive property of transforming according to specific unitary representations of the \Poincare group of special relativity, we have achieved the unification of incoming, outgoing, and regular fields, enabling the use of the same scalar product in all cases. Additionally, we have shown that incoming, outgoing and regular fields transform identically under Lorentz boosts, and derived the closed form matrix element of the Lorentz boost of the polychromatic T-matrix, providing a solid theoretical foundation for investigating the interaction of electromagnetic fields with relativistically moving objects.

To demonstrate the practical implications of our research, we have conducted numerical computations of the transfer of quantities such as energy and momentum from an electromagnetic pulse to a silicon sphere. These results serve as concrete examples of the effectiveness and applicability of the proposed enhancements to the T-matrix method.

By refining the T-matrix method and expanding its capabilities, we provide researchers with valuable tools for advancements in the field of electromagnetic scattering.

\section{Acknowledgments}
This work was funded by the Deutsche Forschungsgemeinschaft (DFG, German Research Foundation) -- Project-ID 258734477 -- SFB 1173.
\newpage
\appendix 

\section{The representation of the vector potential\label{sec:appA}}
The transverse part of the vector potential determines the transverse electric field
\begin{equation}
	\label{eq:EfromA}
	\bm E(\bm r,t)= -\frac{\partial \bm A^{\perp}(\bm r, t)}{\partial t},
\end{equation}
independently of the gauge \cite[Eq.~B.26]{Cohen1997}. In the wave vector space we have hence $\bar{\bm A}^{\perp}(\bm k) = \frac{-i \bar{\bm E}(\bm k)}{ck}$, and the decomposition equivalent to Eq.~(\ref{eq:decomp}) reads
\eq{
	\label{eq:A}
	\bm A^\perp(\bm r, t) &= \sqrt{\frac{\hbar}{c\epsilon_0}} \frac{1}{\sqrt{2}}\frac{-i }{\sqrt{(2\pi)^3}}\sum_{\lambda=\pm 1} \int \frac{d^3 \bm k}{k} \,  f_\lambda (\bm k)\,  \,\bm e_\lambda(\hat {\bm k})  \,e^{i(\bm k \cdot \bm r - c k t)},
}
where the coefficients of the decomposition $f_\lambda (\bm k)$ are the same as the ones of the corresponding electric field. Since one should keep the invariant measure $\frac{d^3 \bm k}{k}$, the decomposition in Eq.~(\ref{eq:A}) induces a definition of plane waves for the vector potential that differs from one of the electric plane waves by the factor of $ik/c$:
\eq{
\bm Q^{A^\perp}_\lambda(\bm k,\bm r, t)&=  - i \sqrt{\frac{\hbar}{ c \epsilon_0}}\, \frac{1}{\sqrt{2}} \frac{1}{\sqrt{(2\pi)^3}}\, \bm e_\lambda(\hat{\bm k}) e^{- i kc t } e^{i \bm k \cdot \bm r}.
}
Vector potential plane waves obey the same transformation rules Eqs.(\ref{eq:wkt1}-\ref{eq:wkt3}) and Eq.~(\ref{eq:qparity}), with an exception of time reversal, where the difference in the imaginary unit $i$ introduces an extra factor of $(-1)$ to the right hand side of Eq.~(\ref{eq:qtime}). The presence of $\frac{\partial }{\partial t}$ in Eq.~(\ref{eq:EfromA}) already announces this difference in time-reversal transformation properties. Also the factor of $k$ difference exactly compensates for the different way that $\bm A^\perp(\bm r, t)$ and $\bm E(\bm r,t)$ transform under Lorentz boosts, namely as the space component of a four-vector, and as the space-time component of an anti-symmetric tensor, respectively.
\section{Lorentz boosts in the $(\mathbf{r},t)$ representation of fields}\label{app:boosts}
Active Lorentz boosts relativistically describe an object moving with a uniform velocity $\bm v$. A 4-vector in Minkowski space-time is transformed under a Lorentz boosts in the z-direction via 
\eq{
	x^\mu = \begin{pmatrix} c t \\ x^1 \\x^2 \\ x^3 \end{pmatrix}  \mapsto L_z(\xi)^\mu_{\;\nu}\, x^\nu = 
\begin{pmatrix}
\cosh(\xi) & 0 & 0 & \sinh(\xi) \\
0 & 1 & 0 & 0 \\
0 & 0 & 1 & 0 \\
\sinh(\xi) & 0 & 0 & \cosh(\xi) \\
\end{pmatrix}
\begin{pmatrix} ct \\ x^1 \\x^2 \\ x^3 \end{pmatrix},
}
with rapidity $\xi = \tanh(v/c)$.
A Lorentz boost in an arbitrary direction can be written as a composition the boost in the z-direction with spatial rotations:
\eq{
	L_{\hat{\bm n}}(\xi) = R(\phi, \theta, 0) L_z(\xi) R^{-1}(\phi,\theta,0). \label{eq:bdecomp}
}
where the direction of the boost $\hat{\bm n}$ is parametrized by polar angle $\theta=\arccos\left(k_z/\abs{\bm k} \right)$ and azimuthal angle $\phi=\arctantwo\left(k_y,k_x\right)$, and the rotations $R$ are parametrized with Euler angles.

In the specific case of a massless 4-wave vector $k^\mu$ with $k^0 = \abs{\bm k}$ the transformation in the z-direction reads
\eq{
k^\mu  = 
\begin{pmatrix}
\abs{\bm k} \\ k^1 \\ k^2 \\ k^3
\end{pmatrix}
\mapsto
\begin{pmatrix}
\cosh(\xi) \abs{\bm k} + \sinh(\xi)k^3 \\ k^1 \\ k^2 \\ \sinh(\xi)\abs{\bm k} + \cosh(\xi)k^3
\end{pmatrix}.
}
and the transformed angles of the wave vector satisfy
\eq{
	\tilde \phi &= \phi\\
	\cos(\tilde \theta) &= \frac{\cos(\theta)\cosh(\xi) + \sinh(\xi)}{\cosh(\xi) + \cos(\theta)\sinh(\xi)} \\
	\sin(\tilde \theta) &= \frac{\sin(\theta)}{\cosh(\xi) +\cos(\theta) \sinh(\xi)}.
}

An active Lorentz boost transformation of real-valued electromagnetic fields, which we distinguish it from the complex fields by a different font, is defined as \cite[Sec.11.10]{Jackson1998}
\eq{
\widetilde{\bm{\mathcal E}} (\bm r, t)&=\gamma {\bm{\mathcal E}}( \tilde{\bm r}, \tilde t) -  \gamma \bm v \times {\bm{\mathcal B}}( \tilde{\bm r}, \tilde t)  - \frac{\gamma^2 \bm v}{(\gamma+1)c^2} \bm v \cdot {\bm{\mathcal E}} ( \tilde{\bm r}, \tilde t) \label{eq:lbu1}\\
\widetilde{\bm{\mathcal B}} (\bm r, t)&=\gamma {\bm{\mathcal B}}( \tilde{\bm r}, \tilde t) +  \frac{1}{c^2}\gamma \bm v \times {\bm{\mathcal E}}( \tilde{\bm r}, \tilde t)  - \frac{\gamma^2 \bm v}{(\gamma+1)c^2} \bm v \cdot {\bm{\mathcal B}} ( \tilde{\bm r}, \tilde t) \label{eq:lbu2}
}
with inversely transformed space-time point $\begin{pmatrix} c\tilde t \\ \tilde{\bm r} \end{pmatrix} =  L^{-1}(\xi) \begin{pmatrix} ct \\ \bm r \end{pmatrix}$ and $\gamma = (1-v^2/c^2)^{-1/2}$. The passive version of the Lorentz boost, i.e. the boost of the reference frame instead of the field, differs from Eqs.(\ref{eq:lbu1}-\ref{eq:lbu2}) by the substitution $\bm v \rightarrow - \bm v$ and should not be confused with the active version. 

One can show that the corresponding Riemann-Silberstein vectors in Eq.~(\ref{eq:RS}) transform under active Lorentz boosts as 
\eq{
\widetilde{\bm F}_\lambda(\bm r, t) &=\gamma  \bm F_\lambda( \tilde{\bm r}, \tilde t) + \frac{i\lambda \gamma}{c} \bm v \times  \bm F_\lambda( \tilde{\bm r}, \tilde t) - \frac{\gamma^2 \bm v}{(\gamma+1)c^2} \bm v \cdot  \bm F_\lambda( \tilde{\bm r}, \tilde t). \label{eq:boost}
}

\section{Transformation properties of $\ket{\mathbf{k} \lambda} $} \label{sec:Qproperties}
Here we derive the transformation laws for plane waves 
\eq{
	\bm Q_\lambda(\bm k,\bm r, t) := \frac{1}{\sqrt{2}}\frac{1}{\sqrt{(2\pi)^3}}\,\sqrt{\frac{c\hbar}{ \epsilon_0}}\, k  \, \bm e_\lambda(\hat{\bm k}) e^{- i kct} e^{i \bm k \cdot \bm r}
}
under the actions of the full Poincar\'e group. The transformation laws for the general electromagnetic field in the $(\bm r,t)$-representation are well-known, and we show how they can be equivalently formulated in the $(\bm k, \lambda)$-domain as a unitary transformation with respect to the scalar product of Eq.~(\ref{eq:scalarPW}).

\subsection{Translations}
Active spatio-temporal translations of electric field by $a^\mu = (\bm a, a^0)$ are defined with
\eq{
 	\widetilde{\bm E}(\bm r, t) = \bm E(\bm r - \bm a, t - a^0).
}
Then, using Eq.~(\ref{eq:decomp})
\eq{
	\widetilde{\bm E}(\bm r, t) 
	&= \frac{1}{\sqrt{2}}\frac{1}{\sqrt{(2\pi)^3}}\sqrt{\frac{c\hbar}{ \epsilon_0}}\sum_{\lambda=\pm 1} \int \frac{d^3 \bm k}{k} \,  f_\lambda (\bm k) \,k\,\bm e_\lambda(\hat {\bm k})  \,  e^{i k_\mu (x^\mu - a^\mu)}\nonumber \\
	&=  \frac{1}{\sqrt{2}}\frac{1}{\sqrt{(2\pi)^3}}\sqrt{\frac{c\hbar}{ \epsilon_0}}\sum_{\lambda=\pm 1} \int \frac{d^3 \bm k}{k} \,  f_\lambda (\bm k) \,p\,\bm e_\lambda(\hat {\bm k})  \, e^{i k_\mu x^\mu} \, e^{-i k_\mu a^\mu}
}
For the plane wave this implies the transformation law
\eq{
	\bm Q_\lambda (\bm k, \bm r - \bm a, t- a^0) &= \frac{1}{\sqrt{2}}\frac{1}{\sqrt{(2\pi)^3}}\,\sqrt{\frac{c\hbar}{ \epsilon_0}}\, k \, \bm e_\lambda(\hat{\bm k}) e^{i k_\mu (x^\mu - a^\mu)} \nonumber \\
	&= \bm Q_\lambda (\bm k, \bm r, t) e^{-i k_\mu a^\mu}
}
which corresponds to Eq.~(\ref{eq:wkt1}).
\subsection{Rotations}
Active rotations of electric field by Euler angles $(\alpha, \beta, \gamma)$, $R(\alpha, \beta, \gamma) = R_z(\alpha)R_y(\alpha)R_z(\gamma)$ are defined in $(\bm r, t)$-domain as \cite[Sec. 6.10]{Jackson1998}
\eq{
	&\widetilde{\bm E}(\bm r, t) = R(\alpha, \beta, \gamma) \,\bm E( R^{-1}(\alpha, \beta, \gamma) \bm r, t)\label{eq:rot-vecfield}
}
where arguments are transformed inversely with respect to the vectorial part. Here and after we use letter $R$ as an abstract operator to describe rotations, and its action depends on the concrete representation of the element that it acts upon. Vectors in the physical three-dimensional space are rotated according to the usual representation of rotations in space, while the polarization vectors $\bm e_\sigma(\hat{\bm k})$ are rotated via the Wigner matrix $D^1(\alpha, \beta, \gamma)$ \cite[Eq.~(76)]{varshalovich1988}: 
\eq{
	R(\alpha, \beta, \gamma) \bm e_\sigma(\hat{\bm k}) =\sum_{\mu=-1,0,1}  \bm e_\mu(\hat{\bm k})\,  D^1_{\mu \sigma }(\alpha, \beta, \gamma) \label{eq:rot-polariz},
}
a special case of which is
\eq{
R_z(\psi) \bm e_\lambda(\hat{\bm z}) = \bm e_\lambda(\hat{\bm z}) e^{-i\lambda\psi} \label{eq:z-rot-polariz}.
}
\Eq{eq:rot-polariz} also allows to find the polarization vector pointing in a general $(\phi, \theta)$-direction in terms of the helicity basis at $\hat{\bm z}$:
\eq{
\bm e_\lambda(\hat{\bm k}) = R(\phi,\theta,0)\bm e_\lambda(\hat{\bm z}) =\sum_{\mu=-1,0,1}  \bm e_\mu(\hat{\bm z})\,  D^1_{\mu \lambda }(\phi, \theta, 0) \label{eq:polariz-decompoz} .
}
Application of the rotation law \Eq{eq:rot-vecfield} to the electric field decomposition Eq.~(\ref{eq:decomp}) gives
\eq{
	\widetilde{\bm E}(\bm r, t) 
	&= \frac{1}{\sqrt{2}}\frac{1}{\sqrt{(2\pi)^3}}\sqrt{\frac{c\hbar}{ \epsilon_0}}\sum_{\lambda=\pm 1} \int \frac{d^3 \bm k}{k} \,  f_\lambda (\bm k) \,k\, R(\alpha, \beta, \gamma) \bm e_\lambda(\hat {\bm k})  \,  e^{-i kc t} e^{i \bm k \cdot (R^{-1}(\alpha, \beta, \gamma) \bm r)} \label{eq:em-rotd}.
}
First, we rewrite the vectorial part as
\eq{
	R(\alpha,\beta,\gamma)  \bm e_\lambda(\hat{\bm k}) &= R(\alpha,\beta,\gamma)R(\phi,\theta,0)\bm e_\lambda(\hat{\bm z})\nonumber  \\
	&= R(\tilde\phi,\tilde\theta,\psi)\bm e_\lambda(\hat{\bm z}),
}
where the rotation by angles $(\tilde\phi,\tilde\theta,\psi)$ realizes the equivalent action to the of two consecutive rotations: $R(\alpha,\beta,\gamma)R(\phi,\theta,0) = R(\tilde\phi,\tilde\theta,\psi)$. Next, we separate the rotation $R(\tilde\phi,\tilde\theta,\psi) = R(\tilde\phi,\tilde\theta,0)R_z(\psi)$ and use Eqs.~(\ref{eq:z-rot-polariz}-\ref{eq:polariz-decompoz}) to simplify the vectorial part to
\eq{
R(\alpha,\beta,\gamma)  \bm e_\lambda(\hat{\bm k})  &=\bm e_\lambda(\hat{\tilde{\bm k}}) e^{-i\lambda\psi},
}
with $\tilde\theta$ and $\tilde\phi $ being the polar and the azimuthal angles of the rotated wave vector $\tilde{\bm k}$.

Transformed scalar part satisfies 
\eq{
e^{i \bm k \cdot (R^{-1}(\alpha, \beta, \gamma) \bm r)}  = e^{i (R(\alpha, \beta, \gamma) \bm k) \cdot \bm r } = e^{i \tilde{\bm k} \cdot \bm r } ,
}
which together with the fact  $\abs{\bm k} = \abs{\tilde{\bm k}}$ brings \Eq{eq:em-rotd} to 
\eq{
\widetilde{\bm E}(\bm r, t) 
	&= \frac{1}{\sqrt{2}}\frac{1}{\sqrt{(2\pi)^3}}\sqrt{\frac{c\hbar}{ \epsilon_0}}\sum_{\lambda=\pm 1} \int \frac{d^3 \bm k}{k} \,  f_\lambda (\bm k) \, \tilde k \, \bm e_\lambda(\hat{\tilde{\bm k}}) e^{-i\lambda\psi} \, e^{-i \tilde kc t}  e^{i \tilde{\bm k} \cdot \bm r } \nonumber \\
	&= \sum_{\lambda=\pm 1} \int \frac{d^3 \bm k}{k} \,  f_\lambda (\bm k) \,  \bm Q_\lambda (\tilde{\bm k}, \bm r, t) e^{-i\lambda \psi}.
}
The last equation implies the required Eq.~(\ref{eq:wkt2}):
\eq{
	\bm Q_\lambda \big(\bm k,\bm r, t\big) \mapsto \bm Q_\lambda (\tilde{\bm k}, \bm r, t) e^{-i\lambda \psi}.
}

\subsection{Parity}\label{sec:parityPW}
Electric field transforms under parity as \cite[Sec. 6.10]{Jackson1998}
\eq{
	\tilde{\bm E}(\bm r, t) &= -\bm E(-\bm r, t). \label{eq:parity-init}
}
Then, using decomposition Eq.~(\ref{eq:decomp})
\eq{
	\tilde{\bm E}(\bm r, t) 
	&= -\frac{1}{\sqrt{2}}\frac{1}{\sqrt{(2\pi)^3}}\sqrt{\frac{c\hbar}{ \epsilon_0}}\sum_{\lambda=\pm 1} \int \frac{d^3 \bm k}{k} \,  f_\lambda (\bm k) \,k\,\bm e_\lambda(\hat {\bm k})  \,  e^{i \bm k \cdot (- \bm r)} \,e^{- i k c t}.
}
For the plane wave this implies the transformation law
\eq{
	\bm Q_\lambda(\bm k, \bm r, t)    &\mapsto  -\bm Q_\lambda(\bm k, -\bm r , t) \nonumber \\
	&= - \frac{1}{\sqrt{2}}\frac{1}{\sqrt{(2\pi)^3}}\, \sqrt{\frac{c\hbar}{ \epsilon_0}}\,k  \, \bm e_\lambda(\hat{\bm k}) e^{-i kc t} e^{i \bm k \cdot (-\bm r)}.
}
Using the definition Eq.~(\ref{eq:ebasis}) one can directly check that 
\eq{
	-\bm e_{\lambda}(\hat{\bm k}) = \bm e_{-\lambda}(-\hat{\bm k}),
}
hence one arrives at
\eq{
	\bm Q_\lambda(\bm k, \bm r, t)    &\mapsto  \bm Q_{-\lambda}(-\bm k, \bm r , t)
}
which corresponds to Eq.~(\ref{eq:qparity}).
\subsection{Time reversal} \label{sec:time-reversal-pw}
Time reversal of real-valued fields is defined as \cite[Sec. 6.10]{Jackson1998}
\eq{
	\tilde{\bm{\mathcal E}} (\bm r, t) &= \bm{\mathcal E} (\bm r, -t).
}
For the complex representation of electromagnetic field Eq.~(\ref{eq:Fcomplex}) this implies 
\eq{
\bm E(\bm r, t) &= \frac{1}{\sqrt{2\pi}}\int_0^{\infty} dk \, e^{-i k c t}\, \tilde{\bm E}(\bm r, k) \mapsto \frac{1}{\sqrt{2\pi}}\int_0^{\infty} dk \, e^{-i kc t}\, \tilde{\bm E}(\bm r, k)^*.
}
Then, using decomposition Eq.~(\ref{eq:decomp}) the time-reversed field can be written as
\eq{
	\tilde{\bm E}(\bm r, t) &= \frac{1}{\sqrt{2}}\frac{1}{\sqrt{(2\pi)^3}}\sqrt{\frac{c\hbar}{ \epsilon_0}}\sum_{\lambda=\pm1} \int \frac{d^3 \bm k}{k} \, \Big[f_\lambda (\bm k) \,k\,\bm e_\lambda(\hat {\bm k})  
	\, e^{i \bm k \cdot \bm r}\Big]^* e^{- i kc t} \nonumber \\
	 &= \frac{1}{\sqrt{2}}\frac{1}{\sqrt{(2\pi)^3}}\sqrt{\frac{c\hbar}{ \epsilon_0}}\sum_{\lambda=\pm1} \int \frac{d^3 \bm k}{k} \, f^*_\lambda (\bm k) \,k\,\bm e_\lambda(-\hat {\bm k})  
	\, e^{i (-\bm k) \cdot \bm r} e^{- i kc t}.
}
Using Eq.~(\ref{eq:ebasis}) one can straightforwardly check that 
\eq{
	\bm e^*_{\lambda}(\hat{\bm k}) = \bm e_{\lambda}(-\hat{\bm k}),
}
hence the plane waves transform as
\eq{
	\bm Q_\lambda(\bm k, \bm r, t) \mapsto \bm Q_\lambda(-\bm k, \bm r, t).
}
which corresponds to Eq.~(\ref{eq:qtime}). The equivalent transformation of the coefficients $f_\lambda(\bm k)$ involves complex-conjugation
\eq{
	f_\lambda(\bm k) \mapsto f^*_\lambda(-\bm k),
}
which reflects the fact that time reversal is represented anti-unitarily.

\section{Parity and time reversal for $\ket{k j m \lambda}$} \label{app:AMparity}
For completeness, we provide here the transformation laws of the relevant angular momentum basis vector fields 
\eq{
\bm R_{j m\lambda}(k, \bm r, t)  &= \frac{1}{\sqrt{\epsilon_0 \hbar c}} \frac{k \, e^{-ikct}}{\sqrt{\pi}\sqrt{2j+1}}  \sum_{L=j-1}^{j+1} \sqrt{2L + 1}\, i^{L} j_{L}(k r) \, C^{j\lambda}_{L0,1\lambda} \bm Y^L_{j m}(\hat{\bm r}) \label{eq:R-eq}\\
\bm S^+_{j m\lambda}(k, \bm r, t)  &= \frac{1}{\sqrt{\epsilon_0 \hbar c}} \frac{k \, e^{-ikct}}{\sqrt{\pi}\sqrt{2j+1}}  \sum_{L=j-1}^{j+1} \sqrt{2L + 1}\, i^{L} h^{\out}_{L}(k r) \, C^{j\lambda}_{L0,1\lambda} \bm Y^L_{j m}(\hat{\bm r})\label{eq:Sp-eq} \\
\bm S^-_{j m\lambda}(k, \bm r, t)  &= \frac{1}{\sqrt{\epsilon_0 \hbar c}} \frac{k \, e^{-ikct}}{\sqrt{\pi}\sqrt{2j+1}}  \sum_{L=j-1}^{j+1} \sqrt{2L + 1}\, i^{L} h^{\inn}_{L}(k r) \, C^{j\lambda}_{L0,1\lambda} \bm Y^L_{j m}(\hat{\bm r})\label{eq:Sm-eq}
}
under parity and time reversal. Here we additionally distinguish between incoming and outgoing fields which is crucial for time reversal and important for the S-matrix formalism.
\subsection{Parity}
We follow the transformation rule mentioned in Sec.~\ref{sec:parityPW} and first consider the transformation of the vector spherical harmonic part \cite{varshalovich1988} 
\eq{
I_s \bm Y^L_{j m}(\hat{\bm r}) &=  - (-1)^L \bm Y^L_{j m}(\hat{\bm r}).
}
Now, with
\eq{
C^{j\lambda}_{L0, 1\lambda} = C^{j-\lambda}_{L0, 1-\lambda} (-1)^{L+1-j}
}
one gets from Eqs.~(\ref{eq:R-eq}-\ref{eq:Sm-eq})
\eq{
	I_s \bm R_{j m\lambda}(k, \bm r, t) &= (-1)^j \bm R_{j m,-\lambda}(k, \bm r, t)\\
	I_s \bm S^\pm_{j m\lambda}(k, \bm r, t) &= (-1)^j \bm S^\pm_{j m,-\lambda}(k, \bm r, t).
}
\subsection{Time reversal}
Now, considering the time reversal of the general electromagnetic field discussed in Sec.~\ref{sec:time-reversal-pw} we use \cite{varshalovich1988}
\eq{
\bm Y^{L}_{j m}(\hat{\bm r})^* &=  (-1)^{L+1+j + m} \bm Y^{L}_{j, -m}(\hat{\bm r})
}
to write
\eq{
	&\Big[ \sum_{L=j-1}^{j+1} \sqrt{2L + 1}\, i^{L} j_{L}(k r) \, C^{j\lambda}_{L0,1\lambda} \bm Y^L_{j m}(\hat{\bm r}) \Big]^* = \nonumber \\ 
	&= \sum_{L=j-1}^{j+1} \sqrt{2L + 1}\, (-1)^L i^{L} j_{L}(k r) \, C^{j\lambda}_{L0,1\lambda}  (-1)^{L+1+j+m} \bm Y^L_{j m}(\hat{\bm r}) \nonumber \\
	&=  - (-1)^{j+m} \sum_{L=j-1}^{j+1} \sqrt{2L + 1}\, i^{L} j_{L}(k r) \, C^{j\lambda}_{L0,1\lambda}  \bm Y^L_{j, -m}(\hat{\bm r})
}
and similarly
\eq{
	&\Big[ \sum_{L=j-1}^{j+1} \sqrt{2L + 1}\, i^{L} h^{\inout}_{L}(k r) \, C^{j\lambda}_{L0,1\lambda} \bm Y^L_{j m}(\hat{\bm r}) \Big]^* = \nonumber \\ 
	&=  - (-1)^{j+m} \sum_{L=j-1}^{j+1} \sqrt{2L + 1}\, i^{L} h^{\inout}(k r) \, C^{j\lambda}_{L0,1\lambda}  \bm Y^L_{j, - m}(\hat{\bm r}),
}
which for the basis states implies 
\eq{
	I_t \bm R_{j m\lambda}(k, \bm r, t) &= -(-1)^{j+m} \bm R_{j , -m.\lambda}(k, \bm r, t) \label{eq:trev1}\\
	I_t \bm S^+_{j m\lambda}(k, \bm r, t) &= -(-1)^{j+m} \bm S^-_{j , -m, \lambda}(k, \bm r, t) \\
	I_t \bm S^-_{j m\lambda}(k, \bm r, t) &= -(-1)^{j+m} \bm S^+_{j , -m, \lambda}(k, \bm r, t)\label{eq:trev3}.
}
As discussed in Sec.~\ref{sec:time-reversal-pw}, one should also conjugate the coefficients of the field $f_{jm\lambda}(k)$ when performing time reversal of the total field. We also note that transformations Eqs.~(\ref{eq:trev1}-\ref{eq:trev3}) result in an extra minus sign compared to the description with the vector potential in App.~\ref{sec:appA}.

\bibliography{PolyTmat} 
\bibliographystyle{unsrt}

\end{document}